\documentclass[12pt]{article}
\usepackage[margin=1in]{geometry}
\usepackage{amsfonts,amssymb,epsfig,amsmath}
\usepackage{slashed}
\usepackage{color}
\usepackage{hyperref}

\renewcommand{\text}[1]{#1}

\newcommand{\be}{\begin{equation}}
\newcommand{\ee}{\end{equation}}
\newcommand{\ben}{\begin{displaymath}}
\newcommand{\een}{\end{displaymath}}
\newcommand{\bea}{\begin{eqnarray}}
\newcommand{\eea}{\end{eqnarray}}
\newcommand{\bean}{\begin{eqnarray*}}
\newcommand{\eean}{\end{eqnarray*}}
\newcommand{\nn}{\nonumber \\}
\newcommand{\ba}{\begin{array}}
\newcommand{\ea}{\end{array}}
\newcommand{\bi}{\begin{itemize}}
\newcommand{\ei}{\end{itemize}}

\def\g{\gamma}
\def\G{\Gamma}

\def\G{\Gamma}
\def\g{\gamma}
\def\e{\epsilon}

\def\e{\epsilon}

\def\m{\mu}

\DeclareMathOperator{\vol}{vol}

\newcommand{\dd}{\mathrm{d}}
\newcommand{\DD}{\mathrm{D}}

\begin{document}

\begin{titlepage}

\vfill
\begin{flushright}
YITP-SB-1404 \\
DMUS--MP--14/01\\
FPAUO-14/01\\
KUNS-2481
\end{flushright}

\vfill

\begin{center}
   \baselineskip=16pt
   {\Large \bf SUSY properties of warped AdS$_3$}
   \vskip 2cm
     Jaehoon Jeong$^a$,  Eoin \'O Colg\'ain$^{b,c,d}$ \& Kentaroh Yoshida$^e$
       \vskip .6cm
             \begin{small}
               \textit{$^a$Department of Physics, College of Science, Yonsei University, Seoul 120-749, KOREA}
                 \vspace{3mm}
                 
                 \textit{$^b$C.N.Yang Institute for Theoretical Physics, SUNY Stony Brook, NY 11794-3840,USA}
                 \vspace{3mm} 
                 
                 \textit{$^c$Department of Mathematics, University of Surrey, Guildford GU2 7XH, UK}
                 \vspace{3mm} 
                 
                 \textit{$^d$Departamento de F\'isica, 
		 Universidad de Oviedo, 
33007 Oviedo, SPAIN}
                 \vspace{3mm} 
                 
      		 \textit{$^e$Department of Physics, Kyoto University, Kyoto 606-8502, JAPAN}
             \end{small}\\*[.6cm]
\end{center}

\vfill \begin{center} \textbf{Abstract}\end{center} \begin{quote}
We examine supersymmetric properties of null-warped AdS$_3$, or alternatively Schr\"{o}dinger geometries,  dual to putative warped CFTs in two dimensions. We classify super Schr\"{o}dinger subalgebras of the superalgebra psu$(1,1|2) \oplus $psu$(1,1|2)$\,, 
corresponding to the superconformal algebra of the AdS$_3\times$S$^3$ geometry. We comment on geometric realisations and provide a string theory description with enhanced supersymmetry in terms of intersecting D3-branes.  For type IIB supergravity solutions based on $T^{1,1}$, we consider the relationship between five-dimensional Schr\"{o}dinger solutions and their three-dimensional null-warped counterparts, corresponding to R symmetry twists. Finally, we study a family of null-warped AdS$_3$ solutions in a setting where there is an ambiguity over the R symmetry and confirm that, for examples admitting a Kaluza-Klein (KK) reduction to three dimensions, the minimisation of a real superpotential of the three-dimensional gauged supergravity captures the central charge and R symmetry. 
\end{quote} \vfill

\end{titlepage}

\tableofcontents
\setcounter{table}{0}
\setcounter{equation}{0}

\section{Introduction} 
It is a well-known property of black holes that the area of the event horizon encodes the entropy of the black hole \cite{Bekenstein:1972tm,Bardeen:1973gs}. Indeed, for classes of supersymmetric black holes with AdS$_3$ near-horizons, it is a further celebrated result from the string theory literature \cite{Strominger:1996sh, Maldacena:1997de} that a microscopic origin for the entropy can be found in terms of the central charge of the dual two-dimensional conformal field theory (CFT). 

While supersymmetric black holes lead to AdS$_3$ near-horizons with SL($2, \mathbb{R}) \times$ SL($2, \mathbb{R})$ symmetry, one simple generalisation is to consider warped AdS$_3$, where the isometry is broken to SL($2, \mathbb{R}) \times$ U($1$). In general, warped AdS$_3$ near-horizons are fairly ubiquitous, cropping up not only as the near horizon of extremal four-dimensional Kerr black holes \cite{Guica:2008mu}, but also residing as vacua in a host of theories including three-dimensional gravity theories with gravitational Chern-Simons terms \cite{Anninos:2008fx, Nilsson:2013fya, Deger:2013yla} and higher-spins \cite{Gary:2012ms}. In fact, null-warped AdS$_3$ solutions can easily be generated via TsT transformations \cite{Lunin:2005jy} and constitute lower-dimensional analogues of Schr\"{o}dinger geometries \cite{Son:2008ye, Balasubramanian:2008dm} of potential relevance to condensed matter. In addition to theories with a gravitational Chern-Simons term, null-warped AdS$_3$ solutions also appear in Maxwell Chern-Simons theories, which are embeddable in string theory \cite{Detournay:2012dz, Karndumri:2013dca}. 

The dual field theories for warped AdS$_3$ are certainly enigmatic. On one hand,  one can still naively apply the Cardy formula to count the degeneracy of states and reproduce the Bekenstein-Hawking entropy \cite{Strominger:1997eq}, thereby hinting that the dual theory may indeed be a CFT with a second hidden Virasoro algebra\footnote{See \cite{oai:arXiv.org:1111.6978, Guica:2013jza, Hartong:2013cba} for related work on holographic renormalization and comments on emergent symmetries.}. An alternative proposal in the literature is that the algebra corresponding to the theory is a single Virasoro algebra with a U($1$) Kac-Moody algebra \cite{Detournay:2012pc} and the dual theory is a more exotic warped CFT, of which there is no non-trivial example. Separately, it has been argued that such field theories arise as IR limits of non-local dipole-deformed theories \cite{ElShowk:2011cm,Song:2011sr}. 

In this paper, we retrace the fact that black holes are microscopically best understood with supersymmetry and it serves as motivation to study null-warped AdS$_3$ spacetimes exhibiting enhanced supersymmetry, a facet of these spacetimes that has been overlooked to date. Along the way, we will study other supersymmetric properties, and since knowledge about the dual field theory is far from concrete, we will be adopting the standard viewpoint that a dual theory can be defined. 

We recall that null-warped AdS$_3$ is indistinguishable from three-dimensional Schr\"{o}dinger geometries with dynamical exponent $z=2$. Therefore, as has been done in higher dimensions \cite{Horvathy, SY1, SY2, SY3}, we start with a classification of the various ways of embedding Schr\"{o}dinger superalgebras in superconformal algebras in three dimensions.  More concretely, we focus on the superalgebra psu(1,1$|$2) $\oplus$ psu(1,1$|$2), corresponding to the superconformal symmetry of the geometry AdS$_3 \times$ S$^3$. Not surprisingly, we identify similar superalgebras to those based on the $\mathcal{N}=4$ superconformal algebra psu(2,2$|$4) and the supersymmetries arrange themselves into ``kinematical", ``dynamical" and ``superconformal", where the latter are generated by the former through the special conformal transformation. However, in addition, we point out the existence of an exotic superalgebra without kinematical supersymmetries, which has no ``higher-dimensional" counterpart.  

With our classification of superalgebras in hand, it is an obvious line of investigation to establish whether any of them admit a geometric realisation. As we have touched on above, given a Schr\"{o}dinger geometry, the field theory picture is largely unclear\footnote{See \cite{Guica:2010sw,Dobrev:2013kha} for comments on the field theory interpretation of a given Schr\"{o}dinger geometry. Given a distinctive non-relativistic Chern-Simons matter theory \cite{Nakayama:2009cz, Lee:2009mm}, one may also search for the dual geometry \cite{Colgain:2009wm,Ooguri:2009cv,Jeong:2009aa}.}, but finding a corresponding supersymmetric geometry to a superalgebra is a simpler task. By considering an intersecting D3-brane solution, following a prescription given in \cite{Donos:2009zf}, it is a relatively simple task to construct null-warped AdS$_3$ solutions exhibiting enhanced supersymmetry; in this case, six, broken down into two kinematical (spectators), two dynamical and two superconformal. To the extent of our knowledge, this is the first supersymmetric null-warped AdS$_3$ solution preserving superconformal supersymmetries. More interestingly still, within this class, one can identify solutions that are essentially direct products and should permit dimensional reductions to three dimensions. 

As stated, the construction of this new solution essentially parallels the recipe given in \cite{Donos:2009zf} for higher-dimensional analogues. Motivated by this fact, we take time to explore what is the relationship between three-dimensional null-warped AdS$_3$ solutions and higher-dimensional supersymmetric Schr\"{o}dinger geometries constructed in \cite{Donos:2009zf}. Our focus is on examples with enhanced supersymmetry, which are novel in the literature (see also \cite{Ooguri:2009cv,Jeong:2009aa}). Starting from five dimensions, we construct the first explicit example with enhanced supersymmetry based on $T^{1,1}$ and remark that the Schr\"{o}dinger solution, like the example based on $S^5$ \cite{Donos:2009zf}, should also correspond to some vacuum of a yet unidentified five-dimensional theory. We then consider standard twists of the R symmetry direction \cite{Maldacena:2000mw}. Recall that in the absence of Schr\"{o}dinger deformations, such twists lead to a flow from a $\mathcal{N}=1$ SCFT \cite{Klebanov:1998hh} to a two-dimensional superconformal fixed-point with $\mathcal{N} = (0,2)$ supersymmetry.  In the presence of Schr\"{o}dinger deformations, we find that geometries with enhanced supersymmetry can be twisted, but in three dimensions the end result corresponds to a solution generated via TsT, so all superconformal supersymmetries are projected out in the process. We find no remaining signature of enhanced supersymmetry after the twist and argue that it should not be expected by considering the projection conditions on the Killing spinors. 

To date, there have been extensive studies of null-warped spacetimes in the context of the well-known AdS$_3 \times$ S$^3 \times$ CY$_2$ solution of type IIB supergravity, where TsT transformations have been applied to generate a host of solutions \cite{Detournay:2012dz,ElShowk:2011cm, Song:2011sr, Bena:2012wc,  Azeyanagi:2012zd}. Some of the motivation of this current work stems from the need to explore other examples, which have been overlooked. To this end, we recall that the general form of supersymmetric AdS$_3$ solutions to type IIB supergravity is known \cite{Kim:2005ez} and it includes a three-parameter family of solutions  \cite{Benini:2012cz, Benini:2013cda} dual to $\mathcal{N} = (0,2)$ SCFTs in two dimensions. By applying TsT transformations, and avoiding the U($1$) R symmetry, we produce potentially the largest class of supersymmetric null-warped AdS$_3$ solutions constructed to date and show generically that they conform with our expectation that a single kinematical supersymmetry is preserved. 

Having not touched the R symmetry in the process\footnote{Otherwise supersymmetry would break \cite{Bobev:2009mw}.}, we are left with a class of supersymmetric null-warped AdS$_3$ solutions with a U($1$) R symmetry\footnote{By this, we simply mean that the Killing vectors will be charged with respect to this direction.}. It is thus expected that $c$-extremization, a procedure to determine the central charge and exact R symmetry of a $\mathcal{N} = (0,2)$ SCFT in two dimensions \cite{Benini:2012cz,Benini:2013cda}, or more precisely, its supergravity dual formulation \cite{Karndumri:2013iqa}, can also be applied here. We recall that Ref. \cite{Karndumri:2013iqa} recasts $c$-extremization in the language of three-dimensional $\mathcal{N} =2$ gauged supergravity and identifies the inverse of the T tensor as the trial central charge. Since the T tensor is built from the embedding tensor, which also appears in the Chern-Simons terms of the supergravity and determines the isometries being gauged, one has a direct relationship between the topological terms and the central charge. Similar conclusions follow from studying 't Hooft anomalies \cite{Benini:2012cz,Benini:2013cda}. 

Starting from the large class of null-warped AdS$_3$ solutions we generate, we find it is possible to preserve supersymmetry, while at the same time ensuring that the higher-dimensional solutions that we generate can also be described in the three-dimensional language. In the process, we identify a two-parameter family of three-dimensional null-warped AdS$_3$ solutions, where the TsT deformations correspond to massive vectors and do not contribute to the Chern-Simons terms of the $\mathcal{N} =2$ sub-sector of the theory. As such, the T tensor is not affected by the TsT transformation and this suggests that $c$-extremization may be immediately generalised to include warped AdS$_3$. It would be interesting if a supporting picture based on anomalies could be established for the dual (warped) CFTs.

The structure of the paper is as follows. In section \ref{sec:Sch}, we study  superalgebras with SL($2,\mathbb{R}) \times$ U($1)$ symmetry, typically referred to as super Schr\"{o}dinger algebras in the literature. In section \ref{sec:D3}, we provide a geometric realisation of one of these algebras and in the process construct the first example of a null-warped AdS$_3$ solution to string theory that exhibits supersymmetry enhancement. Later, in section \ref{sec:twist}, we address the relationship of such solutions to higher-dimensional counterparts corresponding to deformations of AdS$_5$. In section \ref{sec:TsT}, focussing on a three-parameter family of supersymmetric AdS$_3$ vacua, we construct various null-warped AdS$_3$ solutions via TsT transformation and comment on supersymmetry. Finally, in section \ref{sec:gen_case}, we show that a sub-class of the generated solutions can be consistently dimensionally reduced to three dimensions, where the TsT deformations give rise to massive vectors and we comment on the implications for $c$-extremization. In section \ref{sec:conclusions} we conclude and various technical details are housed in the appendix. 

\section{Schr\"{o}dinger superalgebra}
\setcounter{equation}{0}
\label{sec:Sch}

In this section we will give a short summary of super Schr\"odinger algebras relevant to the present analysis, and for the details, the reader is encouraged to consult Appendix A. 

Our starting point is the Lie superalgebra psu$(1,1|2)\oplus$psu$(1,1|2)$\,, 
corresponding to the superconformal algebra of the AdS$_3\times$S$^3$ geometry. 
This superalgebra contains 16 supercharges and 
the related super Schr\"odinger algebras are obtained as subalgebras of the superalgebra 
with the help of projection operators. The basic strategy is the same as in \cite{SY1,SY2,SY3}. 
The bosonic part is universally given by SL$(2,\mathbb{R})\times$U(1)\,. 
The distinguished one is the maximally supersymmetric Schr\"odinger algebra, 
which preserves 4 dynamical, 4 conformal and 4 kinematical supercharges.  
The corresponding geometry is simply the light-like compactification of AdS$_3\times$S$^3$. To appreciate this fact, we recall that the AdS Killing spinors have an $x^-$ dependence, which only drops out when the constant superconformal Killing spinor, $\psi_0$, satisfies $\gamma^+ \psi_0 =0$, thus killing half the superconformal Killing spinors. This leaves twelve. 

It is also possible to obtain less supersymmetric Schr\"{o}dinger superalgebras. 
We present an example that preserves 2 dynamical, 2 conformal and 2 kinematical supercharges with the original SU($2)_L \times$ SU($2)_R$ R symmetry broken to U($1)_{L} \times$ U($1)_{R}$. Later, in section \ref{sec:D3}, we show how such symmetries can be encoded geometrically in a Schr\"{o}dinger deformation of AdS$_3 \times$ S$^3$.   
Finally, in contrast to higher dimensional cases \cite{SY1,SY2,SY3}, we note that some curious structure of 
supercharges is possible in the present case, due to the low-dimensionality. 
As an example, we identify an algebra with 2 dynamical and 2 conformal supercharges, which is closed, without kinematical ones. 
For these less supersymmetric Schr\"odinger algebras, the corresponding geometries are not obvious, but we shouldn't discourage the hunt and it would be satisfying if associated gravitational solutions for each superalgebra could be found.

 
\section{Intersecting D3-branes}  
\setcounter{equation}{0}
\label{sec:D3}
Recall that AdS spacetimes preserve both Poincar\'e and superconformal supersymmetries. As is well appreciated at this stage, Schr\"{o}dinger solutions generated via TsT transformations typically only preserve Poincar\'e supersymmetries, commonly referred to as kinematical supersymmetries. However, with due care it is also possible to find deformations of AdS geometries where extra Poincar\'e Killing spinors, called dynamical supersymmetries, are preserved. In turn these new dynamical supersymmetries generate superconformal Killing spinors, providing a geometric realisation of a super Schr\"{o}dinger algebra. A host of such geometries have been found by considering deformations of known AdS$_5$ solutions to ten and eleven-dimensional supergravity \cite{Ooguri:2009cv, Jeong:2009aa, Donos:2009zf}.  

Here we focus on the analysis presented in \cite{Donos:2009zf} illustrating how five-dimensional Schr\"{o}dinger solutions Sch$_5$ with Sasaki-Einstein manifolds preserving dynamical supersymmetries can be be constructed. Up to a small modification, the same analysis may also be exploited to find similar geometries based on well-known AdS$_3 \times$ S$^3 \times$ CY$_2$ geometries of type IIB supergravity. Here we illustrate the method and refer the reader to \cite{Donos:2009zf} for a more thorough treatment. We believe that this is the first example of a null-warped AdS$_3$ geometry with enhanced supersymmetry. 

We consider the Ansatz
\bea
\dd s^2 &=& r^2 [ 2 \dd x^+ \dd x^- + f (\dd x^+)^2 ] + \frac{1}{r^2} \dd s^2 (\mathbb{R}^4) + \dd s^2 (CY_2), \nn
F_5 &=& 2 (1+ *_{10}) \, r \dd x^+ \wedge \dd x^- \wedge \dd r \wedge J, \nn
G_3 &=&  \dd x^+ \wedge W, 
\eea
where $f$ and $W$ are respectively a scalar and a  complex two-form defined on $\mathbb{R}^4$.  To recover the usual form of the original AdS$_3$ solution when $f=W =0$, we can simply write $\mathbb{R}^4$ as 
\be
\dd s^2 (\mathbb{R}^4) = \dd r^2 + r^2 \dd s^2(S^3). 
\ee

The deformations satisfy the following equations of motion:
\bea
\label{EOMS}
\nabla^2_4 f + |W|^2 &=&0, \nn
\dd W = \dd *_4 W &=& 0, 
\eea
where $*_4$ refers to Hodge duality with respect to $\mathbb{R}^4$ and $|W|^2 = \frac{1}{2} W_{ab} W^{*ab}$. Demanding invariance under the Schr\"{o}dinger algebra, we will be concerned with a deformation $W$, which may be written as 
\be
W = \dd (r^2 \sigma), 
\ee
where $\sigma$ is a complex one-form.

The supersymmetry analysis largely parallels that presented in \cite{Donos:2009zf}. Given the distinct lack of spatial directions for warped AdS$_3$, some notable differences arise, which we comment on in Appendix C. The most striking departure is that we appear to have more freedom and the projection condition $\G^+ \eta_+ = 0$, which is a direct consequence of the above Ansatz and the Killing spinor equations in the higher-dimensional case, does not follow immediately here. However, once one imposes this condition, an analogous solution can be found, which we reproduce here. We remark that the extra freedom we notice here in the Killing spinor equations may be a signature of the presence of solutions corresponding to the exotic superalgebras that exist in three dimensions. 

For $z=2$ and $\G^+ \eta_+ = 0$, as shown in the appendix, the general form of the Killing spinor can be written in terms of Poincar\'e and superconformal Killing spinors, $\epsilon = \epsilon_P+ \epsilon_S$, where 
\bea
\label{Poincare} \e_P &=& r^{\frac{1}{2}} \eta_- - \tfrac{1}{8} r^{\frac{3}{2}} \G^+ \slashed{W} \G^r \eta_-^*, \\
\label{Superconformal} \e_S &=& r^{-\frac{1}{2}} (\G^r - r x^+ \G_+) \eta_+ - \tfrac{1}{4} x^+ r^{\frac{3}{2}} \slashed{W} \G^r \eta_+^*. 
\eea
The spinors $\eta_{\pm}$ only depend on the $\mathbb{R}^4$ coordinates and satisfy the following conditions: 
\bea
\label{cond_summary} \G^+ \eta_+ &=& 0, \nn
\nabla_a^4 \eta_{\pm} &=& 0, \nn
\left(\slashed{\partial} f - \tfrac{1}{8} r \slashed{W} \slashed{W}^* \G^r \right) \eta_+ &=& 0, \nn
\G^+ \left(\slashed{\partial} f - \tfrac{1}{8} r \slashed{W} \slashed{W}^* \G^r \right) \eta_- &=& 0, \nn
\slashed{W} \eta_+ = \G^+ \slashed{W} \eta_- = \slashed{W}^* \eta_- &=& 0. 
\eea
\subsection*{Example}
In order to produce an example, we further decompose the ten-dimensional spacetime into a $(6,4)$-split by writing the gamma matrices as 
\bea
\G_{\mu} &=& \rho_{\mu} \otimes \mathbf{1}_4, \quad \phantom{x} \mu = +, -, 6, 7, 8, 9, \nn
\G_a &=& \rho_{(7)} \otimes \g_a, \quad a = 2, 3, 4, 5, 
\eea
where we have defined $\rho_{(7)} = \rho^+ \rho^- \rho^6 \rho^7 \rho^8 \rho^9$. We further decompose the Killing spinors 
\be
\eta_{\pm} = \xi_{\pm} \otimes \zeta 
\ee
where $\rho^{6789} \xi_{\pm} = - \xi_{\pm}$, $i \rho^{+-67} \xi_{\pm} = \pm \xi_{\pm}$ and $\zeta$ is a (covariantly) constant spinor on $\mathbb{R}^4$ with definite chirality. With this decomposition, the above conditions (\ref{cond_summary}) are satisfied provided 
\bea
\label{condition1} \left(\slashed{\partial} f - \tfrac{1}{8} r \slashed{W} \slashed{W}^* \G^r \right) \zeta &=& 0, \\
\label{condition2} \slashed{W} \zeta = \slashed{W}^* \zeta &=& 0. 
\eea

$\mathbb{R}^4$, and more generally $CY_2$, has a covariantly constant, positive chirality spinor $\zeta$, where the K\"{a}hler form may be written as 
\be
\mathcal{J}_{ab} = i \bar{\zeta}^{\dagger} \g_{ab} \zeta. 
\ee
Moreover, we also have 
\be
\gamma_a \zeta = i \mathcal{J}_{a}^{~b} \g_b \zeta. 
\ee
In terms of holomorphic coordinates, we then have $\gamma^{\bar{\mu}} \zeta =0$, meaning that we can satisfy both conditions (\ref{condition2}) when $W$ is of type $(1,1)$ and primitive \cite{Donos:2009zf}.

This leaves us the task of finding a solution to (\ref{condition1}). Luckily, this has already been executed in \cite{Donos:2009zf} and we can simply quote the essential results. Introducing the one-form 
\be
S \equiv r^2 \sigma, 
\ee
such that $W = \dd S$, $f$ can be solved in terms of $S$ as 
\be
f = - 2 |S^{(0,1)}|^2, 
\ee
where we have isolated the $(0,1)$-component of $S$, 
\be
S_{a}^{(0,1)} \equiv \tfrac{1}{2} \left( S_a + i \mathcal{J}_{a}^{~b} S_b \right).  
\ee
With this result in hand, we can now produce an explicit example. We let $(z_1, z_2)$ denote complex coordinates on $\mathbb{R}^4$ and take 
\be
W = c_1 \dd \bar{z}_2 \wedge \dd z_1 + c_2 \dd \bar{z}_1 \wedge \dd z_2, 
\ee
where $c_i$ are complex constants. One can then work out $S$
\be
2 S = c_1 (\bar{z}_2 \dd z_1 - z_1 \dd \bar{z}_2) + c_2 (\bar{z}_1 \dd z_2 - z_2 \dd \bar{z}_1)
\ee
and its $(0,1)$-component 
\be
2 S^{(0,1)} = - c_1 z_1 \dd \bar{z}_2 - c_2 z_2  \dd \bar{z}_1. 
\ee
This allows one to determine $f$ 
\be
f = - |c_1 z_1|^2 - |c_2 z_2|^2. 
\ee

Writing $\mathbb{R}^4$ in complex coordinates as 
\be
z_1 = r \cos  \tfrac{\theta}{2}  e^{\frac{i}{2} (\psi -\phi)}, \quad z_2 = r \sin  \tfrac{\theta}{2} e^{\frac{i}{2} (\psi +\phi)}, 
\ee
we can bring the metric on $\mathbb{R}^4$ to the form 
\be
\dd s^2 (\mathbb{R}^4) = \dd r^2 + \tfrac{1}{4} r^2 \left[ \dd \theta^2 + \sin^2 \theta \dd \phi^2 + (\dd \psi - \cos \theta \dd \phi)^2 \right], 
\ee
where $S^3$ is written as a Hopf-fibration. Written this way, the metric has the following Killing vectors
\bea
K_1 &=& \cot \theta \cos \psi \partial_{\psi}  + \sin \psi \partial_{\theta} + \frac{\cos \psi}{\sin \theta} \partial_{\phi}, \nn
K_2 &=& \cot \theta \sin \psi \partial_{\psi}  - \cos \psi \partial_{\theta} + \frac{\sin \psi}{\sin \theta} \partial_{\phi}, \nn
K_3 &=& \partial_{\psi}, \quad K_4 = \partial_{\phi}, 
\eea
which correspond to the symmetry $SU(2) \times U(1)$. $K_i$, $i=1, 2, 3$ correspond to the usual left-invariant vector fields and $K_4$ is the additional commuting U($1$). 

We can now work out an explicit expression for $f$ and $W$ immediately above: 
\bea
f &=&  -r^2 \left( |c_1|^2 \cos^2 \tfrac{\theta}{2} + |c_2|^2 \sin^2 \tfrac{\theta}{2} \right)  ,  \nn
W &=& - \tfrac{1}{2} r  (c_1 e^{-i \phi} - c_2 e^{i \phi} ) \dd r \wedge \dd \theta + \tfrac{i}{2} r  \sin \theta (c_1 e^{-i \phi} + c_2 e^{i \phi}) \dd r \wedge \dd \psi \nn
&+&  \tfrac{i}{4} r^2 \cos \theta (c_1 e^{-i \phi} + c_2 e^{i \phi}) \dd \theta \wedge \dd \psi - \tfrac{i}{4} r^2 (c_1 e^{-i \phi} + c_2 e^{i \phi}) \dd \theta \wedge \dd \phi \nn
&+& \tfrac{1}{4} r^2 \sin \theta \dd \phi \wedge \dd \psi (c_1 e^{-i \phi} - c_2 e^{i \phi}). 
\eea
One notes that $\partial_{\psi}$ is still an isometry of the solution and the Lie derivative of $f$ and $W$ with respect to $\partial_{\psi}$ are zero, 
\be
\mathcal{L}_{\partial_{\psi}} f = \mathcal{L}_{\partial_{\psi}} W = 0. 
\ee 
We conclude that the R symmetry is broken to U($1$) by the deformation. It can be checked that the equations of motion (\ref{EOMS}) are satisfied, ensuring that we have a valid supergravity solution. As a special case, we can set $c_1 =c_2 = c$, so that$f$ takes the simple form
\bea
f &=& - |c|^2 r^2. 
\eea

We can also consider another deformation, namely 
\be
W = c (\dd \bar{z}_1 \wedge \dd z_1 - \dd  \bar{z}_2 \wedge \dd z_2), 
\ee
where $c$ is again a complex constant. Again, one can work out $f$ using the prescription above, 
\be
f = - |c|^2 r^2, 
\ee
which is independent of the angular variables. In terms of angular variables, $W$ may be expressed as 
\be
W = i c \, r  \cos \theta  \dd r \wedge \dd \psi - i c \, r \dd r \wedge \dd \phi - \tfrac{i}{2} c \, r^2 \sin \theta \dd \theta \wedge \dd \psi.  
\ee
The Lie derivative of $W$ with respect to both $\partial_{\phi}$ and $\partial_{\psi}$ is now zero, meaning the solution exhibits U($1) \times$ U($1)$ symmetry, in line with the corresponding superalgebra we have noted earlier. 

Note that in various cases above all dependence on the internal $S^3 \times CY_2$ has dropped out from the warped AdS$_3$ metric. This means that there is some three-dimensional theory with this solution. 
 
\subsection*{Superconformal supersymmetries}
We now review some features of the superisometry algebra. Since we do not have any spatial directions in our Schr\"{o}dinger spacetime, the Killing vectors leaving the solution invariant simply correspond to the Hamiltonian $H$, the number operator $N$, the dilatation operator $D$, the generator of special conformal transformations $C$ and, finally, the Killing vectors corresponding to the preserved R symmetry. These may be expressed as follows: 
\bea
H &=& \partial_{+}, \nn
M &=& \partial_{-}, \nn
D &=& r \partial_{r} - 2 x^+ \partial_{+}, \nn
C &=&  (x^+)^2 \partial_+ -  \tfrac{1}{2} r^{-2} \partial_- - x^+ r \partial_r, \nn
R &=& \partial_{\psi}, ~~\partial_{\phi}.
\eea
The generators $H$, $D$ and $C$ satisfy the following commutation relations 
\be
\label{commutators}
[D, H] = + 2 H, \quad [D,C] =-2C, \quad [H,C] = -D. 
\ee
Since $M$ commutes with all other generators, it constitutes a U($1$) and together these symmetries form the expected SL($2, \mathbb{R}) \times$ U($1)$ isometry of null-warped AdS$_3$. 

To confirm that the extra Poincar\'e supersymmetries not annihilated by $\G^+$ generate additional superconformal supersymmetries, we can make use of the spinorial Lie derivative \cite{Kosmann, FigueroaO'Farrill:1999va}
\be
\mathcal{L}_C \e_P \equiv C^M \nabla_M \e_P + \tfrac{1}{8} \dd C_{MN} \G^{MN} \e_P, 
\ee
where $C$ corresponds to a Killing vector, in this particular case of interest, the special conformal Killing vector. Making use of (\ref{psi_plus_eq3}), a calculation shows that 
\be
\mathcal{L}_C \e_P = \tfrac{1}{2} (r^{-1} \G_r - x^+ \G_+ ) \G^+ \e_P - \tfrac{1}{8} r x^+ \slashed{W} \G_r \G^+ \epsilon^*_P, 
\ee
which if one neglects spatial coordinates that no longer exist, the expression above is the same as \cite{Donos:2009zf}\footnote{There is also a factor of $-\frac{1}{2}$ that can be traced to (\ref{commutators}).}. We see that the kinematical supersymmetries, namely those which satisfy $\G^+ \e_P = 0$, cannot generate superconformal Killing spinors. Substituting the expression for the Poincar\'e Killing spinors (\ref{Poincare}), we see directly that 
\be
\mathcal{L}_{C} \epsilon_P =  \tfrac{1}{2} \e_S, 
\ee
once one identifies $\eta_+ = \G^+ \eta_-$. Therefore, we conclude, in line with our expectations,  that the special conformal transformations generate superconformal Killing spinors by acting on the dynamical Killing spinors.

\section{Twist of Sch$_5 \times T^{1,1}$}
\label{sec:twist}
\setcounter{equation}{0}

Maldacena \& Nu\~nez illustrated how one can twist $\mathcal{N} =4$ super Yang-Mills by putting it on a Riemann surface and allowing it flow to a superconformal fixed point in two dimensions \cite{Maldacena:2000mw}. More generally, this procedure can be applied to any four-dimensional $\mathcal{N}=1$ SCFT with a U($1$) R symmetry, with the twist breaking supersymmetry by a half. Here, we  hope to explore what happens when we apply the same procedure to supersymmetric non-relativistic theories with Schr\"{o}dinger symmetry. 

Central to this approach is the key observation that the Schr\"{o}dinger group embeds in the conformal group \cite{Son:2008ye,Balasubramanian:2008dm} in one dimension higher, and it is a well-known fact that the latter maps to the symmetries of AdS. Thus, taking this connection at face value, by deforming AdS$_5$ so that it exhibits Schr\"{o}dinger symmetry, one may hope to capture qualitative features of Schr\"{o}dinger-invariant Chern-Simons matter systems in three dimensions. A collection of potentially relevant field theories can be found in the literature \cite{ Nakayama:2009cz, Lee:2009mm, Nakayama:2008qz, Nakayama:2008td}. 

We begin by reviewing the Maldacena-Nu\~nez procedure in the context of a generic AdS$_5$ solution based on Sasaki-Einstein 
\bea
\dd s^2 &=& \dd s^2(AdS_5) + \dd s^2 (KE_4) + (\dd \psi + P)^2, \nn
F_5 &=&2 \, (1+*_{10}) \left[ J \wedge J \wedge (\dd \psi +P) \right], 
\eea
where $\dd P = 2 J$ and $J$ is the K\"{a}hler-form for the four-dimensional K\"ahler-Einstein metric $KE_4$. The twisting  \cite{Maldacena:2000mw} then leads to an AdS$_3$ solution preserving four supersymmetries: 
\bea
\label{sch_met} \dd s^2 &=& \tfrac{4}{9}\dd s^2(AdS_3) + \tfrac{1}{3} \dd s^2 (H^2) + \dd s^2 (KE_4) + (\dd \psi + P +A)^2, \\
\label{five_form} F_5 &=&2 \, (1+*_{10})\left[  J \wedge ( J - \tfrac{1}{2} F) \wedge (\dd \psi +P +A) \right], 
\eea
Note that to perform the twist, we have simply introduced a gauge field $A$, with field strength $F=\dd A = - \tfrac{1}{3} \vol(H^2)$, so that the gauge field cancels the contribution to the spin connection of the hyperbolic space $H^2$ \footnote{The origin of the various factors and the fact that the Riemann surface must be $H^2$ can all be traced to (\ref{warp_gf}) and (\ref{susy_cond}). While both of these apply strictly to a KK reduction on $S^5$, provided one sets $F^i = F, a_i = a$ and $X_i =1$, further truncating to minimal five-dimensional gauged supergravity in the process, it is known that $S^5$ can be replaced with any generic Sasaki-Einstein space \cite{Gauntlett:2007ma}. Thus, demanding the warp factor $e^{2g}$ is positive, we have $a>0$ and the fact that the Riemann surface is negatively curved follows from (\ref{susy_cond}). Choosing the $H^2$ to have unit radius, we arrive at the factors quoted above. }.

Now, we want to repeat the process, but replace asymptotic AdS$_5$ with the asymptotically Sch$_5$ metric 
\be
\label{sch_flow}
\dd s^2 (\textrm{Sch}_5) = e^{2g(r)} [2 \dd x^+ \dd x^- + f (\dd x^+)^2 + \dd r^2] + e^{2 h(r)} \dd s^2(H^2), 
\ee
while incorporating an appropriate three-form flux deformation to support Schr\"{o}dinger symmetry. Up to the presence of the $g_{++}$ term in the metric, $f$, this is simply the usual Maldacena-Nu\~nez Ansatz. As before, $f$ in general depends on the radial direction $r$ and the internal coordinates. 

\subsection*{Review of $T^{1,1}$} 

For concreteness, we will illustrate this procedure using Schr\"{o}dinger deformations based on $T^{1,1}$. Before proceeding, we review some salient details. $T^{1,1}$ is most easily defined in terms of its Calabi-Yau cone, or ``conifold" \cite{ Klebanov:1998hh, Candelas:1989js}, which can be described by the quadric in $\mathbb{C}^4$
\be
\sum_{A=1}^4 w_A^2 = 0. 
\ee
The complex coordinates $w_A$ transform in the four-dimensional representation of SO($4)$ and have ``charge one" relative to a U($1$). When the Calabi-Yau is written in terms of a cone over a  five-dimensional manifold $T^{1,1}$,  together these symmetries encode those of the coset the $[SU(2) \times SU(2)]/U(1)$.

The Calabi-Yau metric on the conifold can be written explicitly in terms of a K\"ahler potential $\mathcal{F}$, 
\be
\label{CY3}
\dd s^2 (CY_3) = \mathcal{F}' \textrm{tr} (\dd W^{\dagger} \dd W) + \mathcal{F}'' | \textrm{tr} ( W^{\dagger} \dd W) |^2, 
\ee
where the complex coordinates we introduced earlier, namely $w_A$, can be expressed in terms of a matrix, $W = \tilde{r} Z$, where  
\be
Z = \left(  \begin{array}{cc} z_1 & z_2 \\ z_3 & z_4 \end{array} \right) = \left(  \begin{array}{cc} -c_1 s_2 e^{\frac{i}{2} (2 \psi- \phi_1 + \phi_2)} & c_1 c_2 e^{\frac{i}{2} ( 2 \psi - \phi_1 - \phi_2)}  \\ - s_1 s_2 e^{\frac{i}{2} (2 \psi + \phi_1 + \phi_2)}  &  s_1 c_2 e^{\frac{i}{2} (2 \psi + \phi_1 - \phi_2)}\end{array} \right), 
\ee
and we have employed the following shorthand notation, $c_i = \cos \frac{\theta_i}{2}, s_i = \sin \frac{\theta_i}{2}$, $i=1, 2$. The tilde on $r$ is introduced for later convenience. $\mathcal{F}$ denotes the K\"ahler potential, which may be written as 
\be
\mathcal{F} = \tfrac{3}{2} \tilde{r}^{\frac{4}{3}}, 
\ee
and primes refer to derivatives with respect to $\tilde{r}^2$. Inserting $\mathcal{F}$ into (\ref{CY3}), while making the following redefinition 
\be
r = \sqrt{\tfrac{3}{2}} \tilde{r}^{\frac{2}{3}}, 
\ee
one can bring the metric on the Calabi-Yau conifold to the expected form of a cone over the Einstein space $T^{1,1}$
\bea
\label{conifold}
\dd s^2(CY_3) &=& \dd r^2 + r^2 \dd s^2 (T^{1,1}), 
\eea
where we can further write the metric on $T^{1,1}$ as\footnote{Signs of $\phi_i$ appear flipped relative to the conventional form.} 
\be
\dd s^2(T^{1,1}) =  \tfrac{1}{6} \sum_{i=1}^2 (\dd \theta_i^2 + \sin^2 \theta_i \dd \phi_i^2 ) + \tfrac{1}{9} (\dd \psi - \sum_{i=1}^2 \cos \theta_i \dd \phi_i)^2. 
\ee

One can determine the K\"ahler form from the K\"ahler potential 
\bea
\mathcal{J} &=& 
r \dd r \wedge \tfrac{1}{3} ( \dd \psi - \sum_{i=1}^2 \cos \theta_i \dd \phi_i) + \tfrac{1}{6} r^2 \sum_{i=1}^2 \sin \theta_i \dd \theta_i \wedge \dd \phi_i. 
\eea

Having introduced the metric, we can now contemplate the twist. Before doing so, we should explicitly write out a Sch$_5$ geometry with enhanced supersymmetry based on $T^{1,1}$ following the prescription for such spacetimes as given in Ref. \cite{Donos:2009zf}. To the extent of our knowledge, this is the first explicit example based on $T^{1,1}$. Up to flips in the signs of the coordinates $\phi_i$, we make use of the fact that the Killing vectors for $T^{1,1}$ are known. Explicitly, the may be expressed as \cite{Pufu:2010ie}
\bea
&&K^i = \partial_{\phi_i}, \nn
&&K^{i+2} = - \cos \phi_i \partial_{\theta_i} + \cot \theta_i \sin \phi_i \partial_{\phi_i} + \textrm{cosec\,} \theta_i \sin \phi_i \partial_{\psi}, \nn
&& K^{i+4} = - \sin \phi_i \partial_{\theta_i} - \cot \theta_i \cos \phi_i \partial_{\phi_i} - \textrm{cosec\,} \theta_i \cos \phi_i  \partial_{\psi}
\eea
where again $i =1, 2$. When multiplied by an $r^2$ factor, the one-forms dual to the above vectors in a frame tailored to the Calabi-Yau are
\bea
&& r^2 K^i = \tfrac{1}{\sqrt{6}} r \sin \theta_i e^{\phi_i} - \tfrac{1}{3} r \cos \theta_i e^{\psi}, \nn 
&& r^2 K^{i+2} = \tfrac{1}{3} r \sin \theta_i \sin \phi_i e^{\psi} - \tfrac{1}{\sqrt{6}} r \cos \phi_i e^{\theta_i} + \tfrac{1}{\sqrt{6}} r \sin \phi_i \cos \theta_i e^{\phi_i}, \nn 
&& r^2 K^{i+4} = -\tfrac{1}{3} r \sin \theta_i \cos \phi_i e^{\psi} - \tfrac{1}{\sqrt{6}} r \sin \phi_i e^{\theta_i} - \tfrac{1}{\sqrt{6}} r \cos \phi_i \cos \theta_i e^{\phi_i}.
\eea
It is possible to check that $\dd (r^2 K^j)$, $j=1,\dots, 6$ are all primitive $(1,1)$-forms. Furthermore, they are closed by construction and one can check that they are also co-closed in line with EOMs presented in \cite{Donos:2009zf}. By projecting the one-forms onto their $(0,1)$-component, as we did earlier in section \ref{sec:D3}, one can determine $f$: 
\bea
f &=& - r^2 \sum_{i=1}^2 \biggl[ |c_i|^2 ( \tfrac{1}{6} \sin^2 \theta_i + \tfrac{1}{9} \cos^2 \theta_i ) + |c_{i+2}|^2 (\sin^2 \phi_i [ \tfrac{1}{9} \sin^2 \theta_i + \tfrac{1}{6} \cos^2 \theta_i]   + \tfrac{1}{6} \cos^2 \phi_i ) \nn &+& |c_{i+4}|^2 (\cos^2 \phi_i [ \tfrac{1}{9} \sin^2 \theta_i  + \tfrac{1}{6}  \cos^2 \theta_i ]  + \tfrac{1}{6} \sin^2 \phi_i )\biggr],  
\eea
where $c_j$ are again arbitrary complex constants appearing with the above one-forms. By choosing these appropriately, e.g. $c_1 = c_3 = c_5 = c$, $c_2 = c_4 = c_6 = 0$, we can find 
\be
f = - \tfrac{4}{9} |c|^2 r^2. 
\ee
It is interesting to note that the $g_{++}$ term of the metric does not depend on the internal coordinates in this case, so the spacetime factorises into a direct product of Sch$_5$ with $T^{1,1}$. This suggests that there is some lower-dimensional theory, a five-dimensional one, which supports Schr\"{o}dinger geometries with enhanced supersymmetry. 

\subsection*{Twist}

Now that we have identified a suitable five-dimensional Schr\"{o}dinger solution, one with enhanced supersymmetry by construction,  we can ask how one performs the twist? To see this, we consider a concrete example with a complex three-form $G$, built from the one-form dual to the Killing vector $\partial_{\phi_1}$. We can simplify further by taking the constant multiplying the Killing vector to be real, in which case we have no RR two-form and simply a NS two-form
\be
B = c(r) \dd x^+ \wedge [- \tfrac{1}{9} \cos \theta_1 \DD \psi + \tfrac{1}{6} \sin^2 \theta_1 \dd \phi_1]
\ee
where we have explicitly written out the dual one-form. Originally, prior to the twist, we have $c \sim r^2$, but since we are allowing the various warp-factors in the metric (\ref{sch_flow}) to depend on $r$, we also have to allow the same freedom here too. Now, as happens in the relativistic case, we will simply gauge the R symmetry direction $\dd \psi \rightarrow \dd \psi - \frac{1}{y} \dd x$, so that the U($1$) is now fibered over the $H^2$, 
\be
\DD \psi = \tfrac{1}{3} \left( \dd \psi - \cos \theta_1 \dd \phi_1 - \cos \theta_2 \dd \phi_2 - \tfrac{1}{y} \dd x \right). 
\ee
When there is no $g_{++}$ term in the metric and $c=0$ this is precisely the deformation required to flow from the AdS$_5$ vacuum to the AdS$_3$ vacuum. After the twist, our first observation is that $H = \dd B$, while closed by construction, may no longer be co-closed, $\dd * H \neq 0$. In particular, one encounters the equation 
\be
[ c' e^{-g+2h} ]' = 8 c\, e^{g+2h}. 
\ee
Here the warp factor $e^{2h}$ drops out as it is only a constant and $e^{2g} = \frac{4}{9} r^{-2}$ can be read off from (\ref{sch_met}). Primes denote derivatives with respect to $r$. For this equation to be satisfied, $c$ has to scale in a fashion uncharacteristic for Schr\"{o}dinger solutions, notably $c \sim r^{\pm\frac{2}{3} \sqrt{2}}$. We can reinstate the expected $r$ dependence by introducing an $F_3$ term of the form 
\be
F_3 = - \tfrac{1}{3} \kappa e^{2h}   r^{-3} \sin \theta_1 \dd x^+ \wedge \dd r \wedge \dd \theta_1,  
\ee
where we have taken $c(r) = \kappa r^{-2}$, where $\kappa$ is a constant. With this choice for $c(r)$, the other flux equations of motion are then satisfied. 

Even before going any further to identify the rest of the solution, we note a distinct similarity to solutions generated via TsT, which we will discuss in the next section\footnote{The metrics are related up to an overall scale, so $\dd s^2_{\textrm{here}} = \frac{2}{3 \sqrt{3}} \dd s^2_{\textrm{section \ref{sec:TsT}}}$. Taking into account this rescaling, the solutions are the same.}. As we discuss in the next section, these solutions only preserve kinematical Poincar\'e supersymmetries and there is no enhancement. 

In fact, we can also see this directly from the projection conditions. In the notation of \cite{Donos:2009zf}, the ten-dimensional Killing spinor $\epsilon$ can be decomposed into eigenspinors of $\gamma_{D3} = i \G^{+-23}$ 
\be
\epsilon = \e_+ + \e_-,  ~~\textrm{where}~~ \gamma_{D3} \e_{\pm} = \pm \e_{\pm} ~~ \&~~\G^+ \epsilon_+ = 0.  
\ee
The twist introduces an additional projection condition $\G^{23} \epsilon = i \epsilon$ \cite{Maldacena:2000mw,Benini:2013cda}, leading to $\G^{+-} \e_+ = - \e_+ \Rightarrow \e_+ = 0$. In other words, the superconformal Killing spinors of the original five-dimensional geometry are projected out. We are left with 
\be
\G^{+-} \e_- = \e_-. 
\ee
One can further check that $\e_-$ satisfies $\G^{+} \e_- = 0$ as a result. From the viewpoint afforded to us here, this naively looks like it preserves two supersymmetries, but we will see explicitly in the next section that only one survives. 

So to close this section, we review what may be taken away. AdS$_5$ and AdS$_3$ vacua, and their dual CFTs, are related via a twisting procedure. For AdS$_5 \times$ SE$_5$ geometries a prescription exists \cite{Donos:2009zf} to deform the geometry and yet preserve 6 supersymmetries, which may be further broken down into 2 kinematical, 2 dynamical and 2 superconformal. In particular, when SE$_5 = T^{1,1}$, we have shown that one can twist Sch$_5$ geometries to get Sch$_3$, but the prize to be paid is that superconformal supersymmetries get broken.

\section{Warped AdS$_3$ via TsT}
\label{sec:TsT}
\setcounter{equation}{0}

So far, we have explored supersymmetric null-warped AdS$_3$ (Schr\"{o}dinger) solutions with enhanced supersymmetry and their relation to higher-dimensional Schr\"{o}dinger counterparts.  To the extent of our knowledge, the example presented in section \ref{sec:D3}  is the first of its kind. Neglecting this isolated example with enhanced supersymmetry, most null-warped examples to date, or more precisely, those embeddable in string theory, have incorporated TsT \cite{Lunin:2005jy} either directly \cite{Detournay:2012dz, ElShowk:2011cm, Song:2011sr,  Bena:2012wc, Azeyanagi:2012zd}, or as the inspiration for an Ansatz \cite{Bobev:2011qx, Kraus:2011pf}. Most commonly, transformations of the well-known AdS$_3 \times$ S$^3 \times$ T$^4$ solution of type IIB supergravity are considered, so in the first part of this section, we address other possibilities. 

To this end, we return to the classification of supersymmetric AdS$_3$ solutions to type IIB supergravity \cite{Kim:2005ez} and focus on the other notable solution \cite{Naka:2002jz, Gauntlett:2006ns} where the geometry is a direct product. As in higher dimensions, we can then generate a null-warped solution via TsT transformation, which also goes by the name null-Melvin twisting. To preserve supersymmetry care should be taken in isolating global U($1$) symmetries and initial examples of this transformation involved R symmetry directions leading to broken supersymmetry \cite{Herzog:2008wg,Maldacena:2008wh,Adams:2008wt}\footnote{See also \cite{Hartnoll:2008rs, Donos:2009en, Bobev:2009zf,Donos:2009xc, O'Colgain:2009yd} for related constructions.}. It was subsequently realised that some supersymmetry could be preserved when R symmetry directions do not feature in the TsT \cite{Bobev:2009mw}.


To simplify the TsT procedure \cite{Lunin:2005jy}, we will work at the level of an Ansatz that covers the solutions of interest to us. We start from a ten-dimensional Ansatz comprising an AdS$_3$ factor and a circle direction parametrised by $\varphi$ 
\bea
\label{Ansatz}
\dd s^2 &=& e^{2A} \left[  -  \frac{f}{r^4} (\dd x^+ )^2 + \frac{1}{r^2} (2 \dd x^+ \dd x^- + \dd r^2)  \right] + g_{ab} \dd x^a \dd x^b + e^{2B} ( \dd \varphi+P)^2, \nn
F_5 &=& \frac{1}{r^3} \dd x^+ \wedge \dd x^- \wedge \dd r  \wedge A_1 \wedge (\dd \varphi+ P)+ G_4 \wedge (\dd \varphi+ P) 
+ e^{-3A-B} *_6 A_1 \nn &+& e^{3A-B} \frac{1}{r^3} \dd x^+ \wedge \dd x^- \wedge \dd r \wedge *_6 G_4, \nn
B &=& \frac{g}{r^2} \dd x^+ \wedge (\dd \varphi + P), 
\eea
where $g_{ab}$ denotes the metric for the remaining six-dimensional space. $A, B, f, g$, $A_1$, $P$ and $G_4$ are respectively 4 scalars, 2 one-forms and a four-form depending on the six-dimensional space. Other fields may be present, but the effect of TsT on these fields is simply two T-dualities, which is the identity, so we omit them. This Ansatz is simply a minimal set of fields that will play a role in the TsT transformation and when $f =g = 0$ we are starting from an AdS$_3$ geometry supported by a self-dual five-form flux. 

Performing a TsT transformation involving a shift in the null-direction, $x^- \rightarrow x^- + \lambda \varphi$, one generates the following solution:
\bea
\label{postTsT}
\dd s^2 &=& e^{2A} \left[  - [f+ \lambda (2 g  + \lambda e^{2A+2B}) ] \frac{1}{r^4}  (\dd x^+)^2 + \frac{1}{r^2} (2 \dd x^+ \dd x^- + \dd r^2) \right] + g_{ab} \dd x^a \dd x^b \nn &+& e^{2B} ( \dd \varphi+P)^2, \nn
F_5 &=& \frac{1}{r^3} \dd x^+ \wedge \dd x^- \wedge \dd r \wedge A_1 \wedge (\dd \varphi+ P)+ G_4 \wedge (\dd \varphi+ P) 
+ e^{-3A-B} *_6 A_1 \nn &+& e^{3A-B} \frac{1}{r^3} \dd x^+ \wedge \dd x^- \wedge \dd r \wedge *_6 G_4, \nn
F_{3} &=& -\lambda \frac{1}{r^3} \dd x^+ \wedge \dd r \wedge A_1, \nn
B_{2} &=& \left[ g +\lambda e^{2A+2B} \right] \frac{1}{r^2} \dd x^+ \wedge (\dd \varphi + P).
\eea
Note, both the dilaton, if it is non-zero, and the five-form flux are unchanged. In contrast to similar TsT transformations performed on Freund-Rubin products, such as AdS$_5 \times$ SE$_5$, where SE$_5$ denotes a Sasaki-Einstein manifold, here an additional three-form RR flux is generated. We now illustrate how more complicated solutions can be gradually built up by performing multiple TsT transformations using our Ansatz (\ref{Ansatz}). 

\subsection*{Warm-up}
\label{sec:warmup}

Here we consider an example of a simple supersymmetric AdS$_3$ geometry \cite{Naka:2002jz}, which up to an overall rescaling is simply the twist of the AdS$_5 \times T^{1,1}$ solution 
\bea
\label{exp_sol}
\dd s^2 &=& \tfrac{2}{\sqrt{3}}  \frac{1}{r^2} \left[ 2 \dd x^+ \dd x^- + \dd r^2  \right] + \tfrac{\sqrt{3}}{2} \dd s^2(H^2) + \tfrac{\sqrt{3}}{4} \sum^2_{i=1} \dd s^2(S^2_{(i)}) 
+ \tfrac{1}{2 \sqrt{3}} \DD \psi^2, \nn
F_5 &=&   \tfrac{8}{3} \frac{1}{r^3} \dd x^+ \wedge \dd x^- \wedge \dd r \wedge [ \vol(H^2) + \tfrac{1}{8} \sum^2_{i=1} \vol(S^2_{(i)}) ]  \nn
&-&   \tfrac{1}{4} \left[ \vol (S^2_{(1)}) \wedge \vol(S^2_{(2)}) + \tfrac{1}{2} \vol(H^2) \wedge \sum^2_{i=1} \vol(S^2_{(i)}) \right] \wedge \DD \psi, 
\eea
where we have defined $
\DD \psi = (\dd \psi- \sum_i \cos \theta_i \dd \phi_i - \frac{\dd x}{y})$. Here the coordinates $(\theta_i \phi_i)$ and $(x,y)$ parametrise the $S^2$'s and $H^2$ respectively. 

The AdS$_3$ spacetime and Riemann surfaces are all canonically normalised to unit radius. More generally, one is free to replace the product $S^2_{(1)} \times S^2_{(2)}$ with any K\"ahler-Einstein four-manifold, but here we just consider the explicit example above. The vector $\partial_{\psi}$ corresponds to the R symmetry direction with all other U($1$)'s being global. It is expected that transformations involving the R symmetry direction will break all supersymmetry, whereas those involving global U($1$)'s will break only the superconformal supersymmetries \cite{Bobev:2009mw}. We will see this explicitly when we come to discuss supersymmetry. 

As stated, our initial goal here is simply to illustrate how one can gradually generate more complicated solutions from simpler ones. We start by considering a TsT transformation involving a shift with respect to $\psi$, $x^- \rightarrow x^- + \lambda_1 \psi$, since the above solution is already in the correct form and direct comparison with (\ref{Ansatz}) is easy. We will show how this fails to preserve supersymmetry later. Comparison leads to the following identifications: 
\bea
e^{2A} &=& \tfrac{2}{\sqrt{3}}, \quad e^{2 B} = \tfrac{1}{2 \sqrt{3}}, \quad 
A_1 = 0, \quad P = - \sum_i \cos \theta_i \dd \phi_i - \tfrac{1}{y} \dd x, 
\eea
which when plugged directly into (\ref{postTsT}), gives us a new solution. Since $A_1$ is zero, no $F_3$ is generated and the only changes will be to the three-dimensional spacetime parametrised by $(x^+, x^-, r)$ and the inclusion of a $B$-field. Relative to (\ref{exp_sol}), the changes are
\bea
\label{sol1}
\dd s^2_3 &=&  \frac{1}{r^2} \left[ 2 \dd x^+ \dd x^- + \dd r^2  \right]  - \tfrac{1}{3} \lambda_1^2 \frac{1}{r^4} (\dd x^+)^2, \nn
B_2 &=& \tfrac{1}{3} \lambda_1 \frac{1}{r^2} \dd x^+ \wedge \DD \psi. 
\eea

As a next step we can consider TsT with respect to $\phi_1$, $x^- \rightarrow x^- + \lambda_2 \phi_1$. To do this one simply has to recast the solution we have just generated so that it resembles (\ref{Ansatz}). After rewriting, comparison again gives the following identifications: 
\bea
e^{2A} &=& \tfrac{2}{\sqrt{3}}, \quad e^{2 B} = \Delta, \quad f = \tfrac{1}{3} \lambda_1^2, \quad g = - \tfrac{1}{3} \lambda_1 \cos \theta_1, \nn
A_1 &=& \tfrac{1}{3} \sin \theta_1 \dd \theta_1, \quad P = - \tfrac{1}{2 \sqrt{3}} \frac{\cos \theta_1}{\Delta} (\DD \psi + \cos \theta_1 \dd \phi_1), 
\eea
where we have defined
\bea
\Delta = \tfrac{\sqrt{3}}{4} \sin^2 \theta_1 + \tfrac{1}{2 \sqrt{3}} \cos^2 \theta_1. 
\eea

The resulting solution can yet again be determined from (\ref{postTsT}), however in contrast to the previous TsT, here we also generate a three-form RR flux. Again the changes relative to the original solution can be encapsulated as follows: 
\bea
\label{sol2}
\dd s^2_3 &=&  \frac{1}{r^2} \left[ 2 \dd x^+ \dd x^- + \dd r^2  \right]  - \left[ \tfrac{1}{3} \lambda_1^2- \tfrac{2 }{3}  \lambda_1 \lambda_2 \cos \theta_1 + \tfrac{2}{\sqrt{3}}  \lambda_2^2 \Delta \right] \frac{1}{r^4} (\dd x^+)^2, \nn
B_2 &=&\tfrac{1}{3} (\lambda_1 - \lambda_2 \cos \theta_1)\frac{1}{r^2} \dd x^+ \wedge \DD \psi + \tfrac{1}{2} \lambda_2 \sin^2 \theta_1\frac{1}{r^2} \dd x^+ \wedge \dd \phi_1, \nn
F_3 &=& -  \lambda_2 \tfrac{1}{3} \sin \theta_1 \frac{1}{r^3} \dd x^+ \wedge \dd r \wedge \dd \theta_1. 
\eea

We can now repeat another time by performing a TsT involving a shift with respect to $x$, $x^- \rightarrow x^- + \lambda_3 x$. The overall final solution takes the explicit form\footnote{In performing all these transformations one could alternatively perform T-dualities on $x^-$ provided one first works with the corresponding finite temperature solution and one then takes the zero temperature limit. We have checked that the result is unchanged. }
\bea
\label{sol3}
\dd s^{2} &=& - \tfrac{2}{\sqrt{3}} \frac{1}{r^{4}} \left[ \tfrac{1}{3} \lambda_1^2 - \tfrac{2}{3}  \lambda_1 \lambda_2 \cos \theta_1 + \tfrac{2}{\sqrt{3}} \lambda_2^2 \Delta+ \tfrac{2}{3} \frac{ (\lambda_1- \cos \theta_1 \lambda_2) \lambda_3}{y} + \tfrac{4}{3} \frac{ \lambda_3^2}{y^2} \right]  (\dd x^+)^{2}\nn
&+& \tfrac{2}{\sqrt{3}} \frac{1}{r^{2}} \left(2 \dd x^+  \dd x^- + \dd r^{2} \right) + \tfrac{\sqrt{3}}{2} \dd s^2(H^2)+ \tfrac{\sqrt{3}}{4} \sum_i \dd s^2 (S^2_{(i)})
+ \tfrac{1}{2\sqrt{3}} \DD \psi^2
\eea
with NS field,
\bea
B_2 &=&  \frac{1}{r^2} \dd x^+ \wedge \left[ \tfrac{1}{3} \left(\lambda_1 - \lambda_2 \cos \theta_1 - \lambda_3 \frac{1}{y} \right) \DD \psi + \tfrac{1}{2} \lambda_2 \sin^2 \theta_1 \dd \phi_1 + \frac{\lambda_3}{y^2} \dd x \right] , 
\eea
and three-form RR flux 
\bea
F_3 &=&  -   \tfrac{1}{3} \lambda_2 \sin \theta_1  \frac{1}{r^3} \dd x^+ \wedge \dd r \wedge \dd \theta_1+  \tfrac{8}{3} \lambda_3 \frac{1}{ y^2}  \frac{1}{r^3} \dd x^+ \wedge \dd r \wedge \dd y. 
\eea
The five-form flux is unchanged from (\ref{exp_sol}). We bring the reader's attention to the fact that the $g_{++}$ term in the metric generically depends on the internal geometry. We now digress a bit to discuss supersymmetry, before repeating for the general case of the three-parameter family of solutions that featured in Ref. \cite{Benini:2013cda} . 

\subsubsection*{Supersymmetry}
Now that we have generated a simple class of explicit  solutions via TsT, we comment on supersymmetry. Since our original AdS$_3$ geometry can be written locally as a U($1$) fibration over a K\"{a}hler-Einstein six-manifold, further broken down into Riemann surfaces, it bears some resemblance to AdS$_5$ solutions based on Sasaki-Einstein five-manifolds (SE$_5$), where one encounters a U($1$) fibration over K\"{a}hler-Einstein four-manifolds. In each case the U($1$) direction corresponds to the R symmetry. We also recall that supersymmetry preserving TsT transformations for AdS$_5 \times$ SE$_5$ geometries have been studied in \cite{Bobev:2009mw}, where it was noted that  a TsT transformation 
breaks supersymmetry from eight Killing spinors to two Killing spinors provided one avoids the R symmetry. By analogy, in the current setting we expect our original four supersymmetries to be broken to a single supersymmetry. We now illustrate that this is indeed the case. 

In support of this claim we now analyse the Killing Spinor equation (KSE) for the generated solutions.  We take our conventions from \cite{Itsios:2012dc}. Since the geometry is originally supported solely by a five-form flux, the dilatino variation is trivially satisfied and the gravitino variation is satisfied by a Killing spinor of the form 
\be
\label{KS}
\eta = e^{\frac{i}{2} \sigma_2 \psi} \tilde{\eta}, 
\ee
where $\tilde{\eta}$ denotes AdS$_3$ Killing spinors and it is subject to the projection conditions:
\be
\label{proj_cond0}
\G^{xy}  i \sigma_2 \tilde{\eta} = \G^{\theta_1 \phi_1} i \sigma_2 \tilde{\eta} = \G^{\theta_2 \phi_2} i \sigma_2 \tilde{\eta} =  - \tilde{\eta}. 
\ee
From the solution to the KSE, we clearly identify $\psi$ as the R symmetry direction and the existence of four supersymmetries follows from the three commuting projection conditions. 

The net effect of the TsT transformation is to deform the AdS$_3$ factor, through the introduction of a $g_{++}$ component for the metric, while at the same time introducing both a NS and RR three-form flux. The dilatino variation is then no longer trivially zero, but since both $F_3$ and $H= \dd B_2$ have components along the null-direction $x^+$, we can set it to zero by imposing 
\be
\label{G+}
\G^+ \eta = 0. 
\ee

Although this only constitutes a single projection condition, this condition breaks all the superconformal Killing spinors and breaks half the Poincar\'e Killing spinors. Recall that for AdS$_3$, Poincar\'e $\eta_P$ and superconformal Killing spinors $\eta_{SC}$ can be written respectively as 
\be
\eta_{P} = r^{-\frac{1}{2}} \eta_{+}, \quad \eta_{SC} = r^{\frac{1}{2}} \eta_- + r^{-\frac{1}{2}} (x^+ \G^+ + x^- \G^-) \eta_{-},
\ee
where $\G^r \eta_{\pm} = \pm \eta_{\pm}$.  Thus $\G^+$ acting on $\eta_P$ breaks the Poincar\'e Killing spinors by a half, while $\G^+$ acting on $\eta_{SC}$ implies $\eta_- =0$, so we have no superconformal Killing spinors. Thus, the TsT transformation preserves a single Poincar\'e Killing spinor. 

In addition, the projection condition (\ref{G+}) implies that 
\be
\label{new_proj}
\G^{ r \psi} i \sigma_2 \eta =   \eta, 
\ee
 through our chirality condition $\G^{+-r xy \theta_1 \phi_1 \theta_2 \phi_2 \psi} \eta = -\eta$. 

We now move onto the gravitino variation. Recall that we have changed the solution by introducing $F_3, H$ and a metric component $g_{++}$, which affects the spin connection. However, if one imposes (\ref{G+}) these additional terms only affect the gravitino variation $\delta \Psi_{+}$. Some of the Killing spinors of the original geometry will survive provided we can eliminate the terms corresponding to $F_3$ and $H$ is the variation $\delta \Psi_{+}$. Neglecting the other terms coming from the original solution, the relevant expressions are as follows: 
\bea
\label{susy_var}
\delta \Psi_{+} &=& \dots -  \tfrac{1}{8} H_{+ \mu \nu} \G^{\mu \nu}  \sigma_3 \eta - \tfrac{1}{48} F_{\mu \nu \rho} \G^{\mu \nu \rho} \G_{+} \sigma_1 \eta + \dots,  
\eea
where we denote omitted terms through dots. Substituting in the expressions from (\ref{sol3}) we get the following 
\bea
\label{grav_+}
\delta \Psi_{+} &=& \dots + \tfrac{1}{4} \left( \tfrac{2}{\sqrt{3}} \right)^{\frac{1}{2}} \frac{\lambda_1}{r} \sigma_1 \left[ \G^{r \psi} i \sigma_2 + 1 \right] \eta \nn  
&-& \tfrac{1}{4} \left( \tfrac{2}{\sqrt{3}} \right)^{\frac{1}{2}} \cos \theta_1 \frac{\lambda_2}{r} \sigma_1 \left[ \G^{r \psi} i \sigma_2 - 1 \right]  \eta - \tfrac{1}{6} (\sqrt{3})^{\tfrac{1}{2}} \sin \theta_1 \frac{\lambda_2}{r} \sigma_3 \left[  \G^{r \psi} i \sigma_2 -1 \right] \eta  \nn &-& \tfrac{1}{4} \left(\tfrac{2}{\sqrt{3}} \right)^{\frac{1}{2}} \frac{\lambda_3}{ry} \sigma_1 \left[ \G^{r \psi} i \sigma_2 -1  \right] \eta - \tfrac{1}{12} (2 \sqrt{3})^{\frac{1}{2}}  \frac{\lambda_3}{ry}  \G^{y \psi} \sigma_3 \left[ \G^{r \psi} i \sigma_2  -  1 \right]  \eta \nn &+& \dots 
\eea
where we have made use of the projection conditions (\ref{proj_cond}). Observe that the $\lambda_1$ term comes with the wrong sign and will break supersymmetry. 

So, to summarise, provided we do not touch the R symmetry direction we expect all the geometries to preserve one supersymmetry. This is consistent with observations made in \cite{Bobev:2009mw}. 

\subsection*{General case}
Having illustrated the procedure for performing multiple TsT transformations on an explicit example, here we switch our attention to a general class of supersymmetric AdS$_3$ solutions to type IIB supergravity parametrised by three parameters $a_i$, $i=1, 2, 3$. These solutions generically possess $U(1)^4$ symmetry, three of which come from the reduction on $S^5$ and the remaining U($1$) corresponds to a symmetry of a Riemann surface comprising part of the solution\footnote{When the Riemann surface is $T^2$, we naturally have an additional U($1$), but regardless of whether we have $H^2, T^2$ or $S^2$, we can always isolate a U($1$) direction.}. From the three U($1$)'s originating from the $S^5$ a particular linear combination, which may be determined either by $c$-extremization \cite{Benini:2012cz,Benini:2013cda,Karndumri:2013iqa} or directly in higher dimensions \cite{Kim:2005ez}, corresponds to the R symmetry, so if we wish to generate supersymmetric solutions via TsT then, without considering other transformations, we can only consider three transformations. 

We immediately review the class of AdS$_3$ solutions as they appeared in Ref. \cite{Benini:2013cda}.  In ten dimensions the original solutions may be expressed as 
\bea
\label{orig_ansatz}
\dd s^2 &=& \Delta^{\frac{1}{2}} \dd s^2_5 + \Delta^{-\frac{1}{2}}  \sum_i X_i^{-1} \left( \dd \mu_i^2 + \mu_i^2 ( \dd \varphi_i + A^i)^2 \right), \\
F_{(5)} &=& (1 + *_{10} ) \sum_i \biggl[ 2 X_i (X_i \mu_i^2 - \Delta ) \vol_5 + \tfrac{1}{2} X_i^{-2} \dd (\mu^2_i) \biggl( ( \dd \varphi_i + A^i) \wedge *_5 F^i + X^i *_5 \dd X^i \biggr)  \biggr], \nonumber
\eea
where $ \Delta = \sum_i X_i \mu_i^2$ and the five-dimensional part can further be written in terms of a genus $\frak{g}$ Riemann surface $\Sigma_{\frak{g}}$: 
\bea
\dd s^2_5 &=& e^{2 f} \dd s^2(AdS_3)+ e^{2 g} \dd s^2 (\Sigma_{\frak{g}}), \nn
F_i &=& - a_i \vol (\Sigma_{\frak{g}} ), 
\eea
where the AdS$_3$ radius is set to one, $\ell = 1$. Closure of $F_i$ demands that $a_i$ are constant. In terms of the scalars $X_i$ (there is a review of $U(1)^3$ gauged supergravity in the appendix), the five-dimensional warp factors may be expressed as 
\be
\label{warp_gf}
e^{2g} = \tfrac{1}{2} (a_1 X_2 + a_2 X_1), \quad e^{f} = {2}{(X_1 +X_2 + X_3)^{-1}}, 
\ee
where ($X_1 X_2 X_3 =1$)
\be
X_1 = \frac{a_1(-a_1 + a_2 + a_3)}{(a_1+a_2-a_3)} X_3, \quad X_2 = \frac{a_2(a_1 - a_2 + a_3)}{(a_1+a_2-a_3)} X_3. 
\ee

For a supersymmetric AdS$_3$ vacuum, one demands \cite{Benini:2012cz,Benini:2013cda}
\be
\label{susy_cond}
a_1 + a_2 + a_3 = - \kappa, 
\ee
where $\kappa$ is the curvature of the Riemann surface $\Sigma_{\frak{g}}$. The canonical Killing vector dual to the R symmetry may be expressed as 
\be
\label{Rsym}
\partial_{\psi} = \sum_i \frac{X_i}{X_1 + X_2 + X_3} \partial_{\varphi_i}. 
\ee



\subsection*{TsT}
In addition to the obvious three $U(1)'s$ parametrised by $\varphi_i$, $i=1, 2, 3$, we can also consider a TsT transformation with respect to the U($1$) on the constant curvature Riemann surface. One can incorporate the three possibilities for constant  curvature Riemann surfaces with the following parametrisation for the space: 
\be
\dd s^2(\Sigma_{\frak{g}}) = \frac{4 ( \dd \rho^2 + \rho^2 \dd \beta^2)}{(1 + \kappa \rho^2)^2}, 
\ee
where $\kappa = -1, 0, 1$ corresponds to the choice of constant curvature. With this choice for the metric on the Riemann surface, the three U($1$) gauge potentials become 
\be
A_i = -  \frac{2 a_i  \rho^2 }{(1+\kappa \rho^2)} \dd \beta, 
\ee
where $\partial_{\beta}$ now corresponds to an additional U($1$) Killing direction with respect to which we can perform TsT. 

We now move onto performing the TsT's. Using our earlier Ansatz (\ref{Ansatz}), we can now perform TsT transformations with respect to $\varphi_i$ and $\beta$ in turn. The end result is 
\bea
\label{gensol1}
\dd s^2 &=& \Delta^{\frac{1}{2}} e^{2 f} \biggl( - e^{2 f} \biggl[ \sum_{i=1}^3 \frac{\mu_i^2}{X_i} \left( \lambda_i -   \frac{ \lambda_4\, 2 a_i \rho^2}{(1+\kappa \rho^2)} \right)^2  \biggr]  \frac{1}{r^4} (\dd x^+)^2 + \frac{1}{r^2} \left[ 2 \dd x^+ \dd x^- + {\dd r^2} \right] \biggr) \nn &+& \Delta^{\frac{1}{2}}  e^{2 g} \dd s^2(\Sigma_{\frak{g}}) + \Delta^{-\frac{1}{2}}  \sum_i X_i^{-1} \left( \dd \mu_i^2 + \mu_i^2 ( \dd \varphi_i + A^i)^2 \right), \nn
B &=& \sum_{i=1}^3 \frac{e^{2 f}}{X_i}   \left({\lambda_i} -  \frac{ \lambda_4\, 2 a_i \rho^2}{(1+\kappa \rho^2)} \right) \frac{1}{r^2} \dd x^+ \wedge  \mu_i^2 (\dd \varphi_i + A^i)    
\eea

As we remarked before, the five-form flux is unchanged, while the accompanying three-form RR flux is 
\bea
\label{gensol2}
F_3 &=& \sum_{i=1}^3 \frac{1}{r^3} \dd x^+ \dd r \, e^{3f-2g}  \biggl[ (\lambda_i -  \frac{\lambda_4 2 a_i \rho^2}{(1+\kappa \rho^2)} ) \frac{a_i}{X_i^2} \mu_i \dd \mu_i  \nn &-& \lambda_4 2 X_i (X_i \mu^2 - \Delta) e^{4g} \frac{4 \rho }{(1+\kappa \rho^2)^2} \dd \rho \biggr]. 
\eea

Given that we started with a three-parameter family of AdS$_3$ solutions and performed four TsT transformations, it is expected that this solution constitutes one of the largest classes of null-warped AdS$_3$ solutions in a string theory context. As we shall see in due course, with a little gymnastics to avoid the R symmetry, one can also find supersymmetric solutions\footnote{A similar calculation appeared recently in \cite{Mallayev:2013pwa}.}. 

We remark that we can extend the above solutions through S-duality transformations. Following each TsT transformation, it is possible to perform a transformation of the form 
\be
\tau \rightarrow -\frac{1}{\tau}, \quad \left( \begin{array}{c} C_2 \\ B_2 \end{array} \right) \rightarrow \left( \begin{array}{c} -B_2 \\ ~C_2 \end{array} \right)
\ee
where $\tau = C_0 + i e^{-\phi} $ combines the axion $C_0$ and the dilaton $\phi$. This transformation, a particular case of the SL($2, \mathbb{R}$) symmetry of type IIB supergravity, leaves the dilaton unchanged and switches the RR and NS two-forms, $C_2$ and $B_2$ respectively. Most importantly, it is known that S-duality simply rotates the Killing spinors \cite{Ortin:1994su}, so supersymmetry is preserved. 

\subsection*{Supersymmetry}
In this section we take a quick look at the supersymmetry. Our approach will be morally the same as in section \ref{sec:warmup}. While the original AdS$_3$ are suitably simple from the five-dimensional perspective, once one considers the full ten-dimensional solution, namely the setting where we can generate new solutions via TsT, the identification of the Killing spinors becomes a difficult task. For this reason we will bypass the analogous step of identifying the Killing spinors, or at the very least, the projection conditions, for the original class of solutions. We will also not consider TsT with respect to the Riemann surface U($1$) Killing vector $\partial_{\beta}$ (we take $\lambda_4 =0$), as it should be clear that this is a global U($1$) and it does not mix with the R symmetry. 

Instead, we will simply impose (\ref{G+}), in the process breaking all but one supersymmetry, before focussing on the deformations from the original geometry. The metric component $g_{++}$ leads to deformations that are projected out of the Killing spinor equations through (\ref{G+}), so we simply have to concentrate on the contribution from the terms in (\ref{susy_var}). Using the orthonormal frame in the appendix (\ref{7d_ortho}), one can identify the corresponding matrix (\ref{susy_matrix}). Then, plugging the matrix into mathematica and evaluating the determinant we find a necessary condition for supersymmetry\footnote{We have used an explicit representation for the gamma matrices presented in Ref. \cite{OColgain:2012ca}.} 
\be
\label{cond_susy}
\sum_{i=1}^3 \lambda_i = 0. 
\ee 
This condition ensures we have at least one zero eigenvalue, and thus, some conserved supersymmetry. It is expected that a single supersymmetry is preserved. We can also check that it is consistent with the simplest case where $X_i = 1$, in which case we see that the TsT vector is not along the R symmetry direction. 

\section{Comments on $c$-extremization}
\setcounter{equation}{0}
\label{sec:gen_case}

We recall that $c$-extremization \cite{Benini:2012cz, Benini:2013cda} is a procedure to extract the central charge and R symmetry for two-dimensional CFTs with $\mathcal{N}= (0,2)$ supersymmetry. A related statement for AdS$_3$ vacua can be found for three-dimensional $\mathcal{N} =2$  gauged supergravities \cite{Karndumri:2013iqa}, where the so-called T tensor\footnote{In general it is a tensor, whereas with $\mathcal{N}=2$ supersymmetry, it is simply a scalar.},  corresponding to a real superpotential, is extremised. Schr\"{o}dinger spacetimes that result from TsT transformations have been related in the literature \cite{ElShowk:2011cm} to so-called null dipole theories \cite{Bergman:2000cw,Bergman:2001rw,Dasgupta:2001zu}. On the assumption that we still have a dual field theory description and some preserved supersymmetry post TsT, it is worth considering if the $c$-extremization procedure outlined in \cite{Karndumri:2013iqa} can also be applied to null-warped AdS$_3$ solutions, where an ambiguity over the exact R symmetry arises. 

Even this modest question, unrelated to any field theory treatment or discussion of associated anomalies, may be difficult to answer in a more general setting than the scope afforded by this work.  One would need to identify a completely generic null-warped AdS$_3$ solution with an ambiguity over the R symmetry. In contrast, in the previous section we have applied TsT with the objective of not breaking supersymmetry. In the process, we have left the R symmetry untouched, so for the class of null-warped AdS$_3$ solutions discussed in the previous section, we expect that the prescription given in \cite{Karndumri:2013iqa} still applies. We will now show how this is the case. 

Indeed, as we shall observe in this section, there is a two-parameter family of supersymmetric solutions that can be KK reduced to three dimensions. From the three-dimensional perspective, the TsT deformations simply source \textit{massive} vector fields and not Chern-Simons terms. Since it is only topological terms that are related to the T tensor, in particular the T tensor of the $\mathcal{N} =2$ sub-sector of the three-dimensional theory, we conclude that the supergravity dual of $c$-extremization can be applied equally well to null-warped AdS$_3$ solutions.

So, the task of the rest of this section is to rewrite the TsT transformations in a three-dimensional fashion. To aid this, we quickly review the three-dimensional set-up in the absence of deformations coming from TsT. As shown in Refs. \cite{Karndumri:2013dca, Karndumri:2013iqa} (see also \cite{Cucu:2003yk} for earlier work), five-dimensional $U(1)^3$ gauged supergravity can be consistently KK reduced on a constant curvature Riemann surface to give $\mathcal{N} =2$ gauged supergravity in three dimensions. The action takes the form 
\bea
\label{Einsteinact}
\mathcal{L}^{(3)} &=&  R *_3 \mathbf{1} - \tfrac{1}{2} \sum_{i=1}^3 \left[ \dd W_i \wedge * \dd W_i + e^{-2 W_i} \DD Y_i
\wedge * \DD Y_i \right]   \nn &+&
4 g^2 \left[
e^{-W_1-W_3} + e^{-W_2  -W_3} + e^{-W_1-W_2} \right]  +  2 \kappa
e^{-W_1 -W_2 -W_3} \nn &-& \tfrac{1}{2} \left[ a_1^2 \,
e^{-2(W_2+W_3)} + a_2^2 \, e^{-2(W_1+W_3)} + a_3^2 \,
e^{-2(W_1+W_2)} \right] \nn &+& a_1 B^2 \wedge \dd B^3 + a_2 B^3 \wedge \dd B^1
+ a_3 B^1 \wedge \dd B^2,  
\eea
where we have made use of the following redefinitions of the scalars 
\bea
W_1 &=& 2 C + \tfrac{1}{\sqrt{6}} \varphi_1 + \tfrac{1}{\sqrt{2}} \varphi_2, ~~W_2 = 2 C + \tfrac{1}{\sqrt{6}} \varphi_1 - \tfrac{1}{\sqrt{2}} \varphi_2, ~~W_3 = 2 C - \tfrac{2}{\sqrt{6}} \varphi_1
\eea
and the covariant derivatives can be defined further as 
\bea
\DD Y_1 &=& \dd Y_1 + a_3 B^2 + a_2 B^3,\nn \DD Y_2 &=& \dd Y_1 + a_1 B^3 + a_3 B^1, \nn
\DD Y_3 &=&  \dd Y_3 + a_1 B^2 +a_2 B^1.  
\eea
In addition to the two original scalars $\varphi_i$ of the five-dimensional gauged supergravity, an extra scalar, $C$, corresponding to the breathing mode of the Riemann surface, arises from the reduction procedure.  As explained in \cite{Karndumri:2013iqa}, the T tensor, which is itself the real superpotential for the scalar potential, encodes information about the exact R symmetry and central charge. 

In this section, we show that there is a consistent KK dimensional reduction, including the TsT transformations, to a three-dimensional theory. In terms of dimensional reductions, type IIB supergravity reduces on $S^5$ to give maximally supersymmetric $SO(6)$ gauged supergravity in five dimensions \cite{s51}. Group theory dictates that we can further truncate to the $U(1)^3$ Cartan subalgebra giving rise to five-dimensional $U(1)^3$ gauged supergravity \cite{Cvetic:1999xp}, which we review in the appendix. Both of these reductions are based on a Freund-Rubin Ansatz and neither the NS nor RR three-form flux feature. As we have seen, TsT takes us out of the class of these reductions, since three-form fluxes are generated. 

Thus, the process of identifying the three-dimensional gauged supergravity corresponding to our TsT transformations would be greatly simplified if it was known how to extend $U(1)^3$ gauged supergravity in five dimensions to include fields coming from the NS and RR three-forms. To the extent of our knowledge, a fitting consistent KK reduction has yet to be identified\footnote{The closely related $SU(2) \times U(1)$ reduction of type IIB has three-form fluxes \cite{oai:arXiv.org:hep-th/9909203}.}. In the absence of the existence of such a reduction, here we will assume that all gauge fields are null\footnote{Given a $p$-form $A$ and $q$-form $B$, we assume $A \wedge * A = A \wedge B = A \wedge *B = B \wedge * B = 0$.} and that the scalars $e^{g}, X_i$ only depend on the radial direction of (null-warped) AdS$_3$. 

We employ the following three-form flux Ansatz, which is motivated by our earlier results on TsT transformations 
\bea
\label{3form_ansatz}
H_3 &=& \sum_{i=1}^3 \biggl[ (\dd b^i - 2 B_1^i) \wedge \mu_i \dd \mu_i \wedge (\dd \varphi_i + A^i) + \mu_i^2 H_2^i \wedge (\dd \varphi_i + A^i)  \nn
&-& \tfrac{1}{2} \mu_i^2 a_i (\dd b^i -2 B^i_1) \wedge \vol(\Sigma_{\frak{g}}) + a_i  B^i_2  \wedge \mu_i \dd \mu_i  \biggr], \nn
F_3 &=& \sum_{i=1}^3 \biggl[ (\dd c^i - 2 C_1^i) \wedge \mu_i \dd \mu_i \wedge (\dd \varphi_i + A^i) + \mu_i^2 G_2^i \wedge (\dd \varphi_i + A^i)  \nn
&-& \tfrac{1}{2} \mu_i^2 a_i (\dd c^i -2 C^i_1) \wedge \vol(\Sigma_{\frak{g}}) + a_i  C^i_2  \wedge \mu_i \dd \mu_i  \biggr], 
\eea
where $H_2^i = \dd B_1^i$, $G_2^i = \dd C_1^i$. In total, we have introduced six scalars $b^i, c^i$, six vectors $B^i_1, C^i_1$ and six two-forms $B^i_2, C^i_2$. In the absence of the dilaton and axion, the Bianchi identities, namely $\dd H_3 = \dd F_3 =0$, are satisfied once the two-forms are closed. The remaining Bianchi, $\dd F_5 = H_3 \wedge F_3$, is unchanged since we have assumed that the fields we have introduced are all null. Our spacetime Ansatz is up to a conformal transformation largely the same as (\ref{orig_ansatz}), which is designed to bring us to Einstein frame in three dimensions
\be
\dd s^2 = \Delta^{\frac{1}{2}}[ e^{-4g} \dd s^2_3 + e^{2g} \dd^2 (\Sigma_{\frak{g}})  ] + \Delta^{-\frac{1}{2}}  \sum_i X_i^{-1} \left[ \dd \mu_i^2 + \mu_i^2 ( \dd \varphi_i + A^i)^2 \right], 
\ee
and the five-form flux is unchanged. Here $\dd s^2_3$ denotes the three-dimensional metric, which will \textit{a priori} depend on the coordinates of the internal space through its $g_{++}$ component. We will comment on this in due course.  

Now, on the assumption that the scalars only depend on the radial direction, one can show that the three-form flux equations of motion imply the following: 
\bea
\label{cond1} B_2^i &=& -\tfrac{1}{2} X_i^{-1} e^{-4g} *_3 (\dd c^i - 2 C_1^i), \quad
C_2^i = \tfrac{1}{2} X_i^{-1} e^{-4g} *_3 (\dd b^i - 2 B_1^i), \\
\label{cond2}0 &=& \dd \left[ e^{4 g} X_i  *_3 H_2^i \right] -\tfrac{1}{2} \frac{e^{-4 g}}{X_i} \left[ a_i^2 + 4 X_i^3 e^{4g} \sum_{j\neq i} X_j \right] * (\dd b^i - 2 B^i_1),  \\
\label{cond3} 0 &=& \dd \left[ e^{4 g} X_i *_3 G_2^i \right] - \tfrac{1}{2} \frac{e^{-4 g}}{X_i} \left[ a_i^2 + 4 X_i^3 e^{4g} \sum_{j\neq i} X_j \right] * (\dd c^i -2 C^i_1),   
\eea
where in the last two equations the $i$ index is not summed. As a result of this exercise, we see that the two-forms $B^i_2, C^i_2$ are not independent and can be eliminated in terms of the scalars and vectors. 

Setting the scalars to zero $b^i = c^i =0$, we can redefine 
\bea
Y^i = - 2 * B_1^i = e^{f +2 g} H_2^i, ~~\mbox{or}~~Y^i = - 2 * C_1^i = e^{f +2 g} G^i_2
\eea
in order to recast (\ref{cond2}) and (\ref{cond3}) so that they take the expected form for a solution exhibiting $z=2$ scaling\footnote{See section 5 of \cite{Karndumri:2013dca}.} 
\be
\dd *_3 Y \pm \frac{z}{\ell} Y = 0, 
\ee
where in the explicit example of interest 
\be
\ell = \frac{2 a_1 a_2 a_3}{2 (a_1^2 a_2^2 + a_3^2 a_1^2 + a_2^2 a_3^2) - a_1^2 - a_2^2 -a_3^2}, 
\ee and $z=2$. 

Observe that we have performed a conformal transformation with the intention of producing a three-dimensional theory in Einstein frame. Therefore, in contrast to the original solutions (\ref{orig_ansatz}), where $\ell =1$, we have the unusual choice of $\ell$ immediately above. 

We can also work out the changes in the Einstein equation. Dropping terms that feature in the original KK reduction, we get 
\bea
\label{Ein0}
\Delta^{-\frac{1}{2}} e^{4g} \bar{R}_{\mu \nu} &=& \cdots + \sum_{i=1}^3 \biggl[\tfrac{1}{2} \Delta^{-\tfrac{1}{2}} e^{8g} \mu_i^2 X_i (H_2^i)^2_{\mu \nu}  + \tfrac{1}{2} \Delta^{-\frac{1}{2}} e^{8g} \mu_i^2 X_i (G_2^i)^2_{\mu \nu}   \nn &+& \tfrac{1}{8} \Delta^{-\frac{1}{2}} X_i^{-1} \left[  \mu_i^2 a_i^2 + 4 X_i^3 e^{4g}  (\Delta - \mu_i^2 X_i) \right] (\dd b^i - 2 B_1^i)^2_{\mu \nu} \nn 
&+& \tfrac{1}{8} \Delta^{-\frac{1}{2}} X_i^{-1} \left[  \mu_i^2 a_i^2 + 4 X_i^3 e^{4g}  (\Delta - \mu_i^2 X_i) \right] (\dd c^i - 2 C_1^i)^2_{\mu \nu}  \biggr] + \cdots
\eea
where we have just focussed on the Einstein equation along the three-dimensional spacetime, since the fluxes we consider are all null and do not affect the other components of the Einstein equation. Here dots denote omitted terms from the original reduction. 

We note at this point that there is an inconsistency between the equations of motion coming from the flux (\ref{cond2}), (\ref{cond3}) and the Einstein equation (\ref{Ein0}), where different factors appear. This is simply highlighting the fact that the solution generated by TsT generically has a $g_{++}$ metric component that depends on the internal geometry. We can make a simple choice based on our earlier TsT result (\ref{gensol1}) to restore consistency. Neglecting supersymmetry for the moment, we can consistently set $b^i = c^i = C_1^i = 0$ and identify the remaining fields as follows 
\be
B_1^i =  \sqrt{\frac{X_1}{X_i}} \mathcal{B}_1 ~~ \Rightarrow ~~ H_2^i =  \sqrt{\frac{X_1}{X_i}} \mathcal{H}_2.
\ee

Setting the scalars $X_i$ to their AdS$_3$ vacuum values, thus allowing us to identify 
\bea
\frac{\left[ a_1^2 + 4 X_1^3 e^{4g} (X_2 + X_3) \right]}{X_1^2} = \frac{ \left[ a_2^2 + 4 X_2^3 e^{4g} (X_1 + X_3) \right] }{X_2^2} = \frac{ \left[ a_3^2 + 4 X_3^3 e^{4g} (X_1 + X_2) \right] }{X_3^2}, 
\eea
the contribution of the vector $\mathcal{B}$ to the action takes the form of a massive vector: 
\be
\delta \mathcal{L}^{(3)} = \tfrac{1}{2} e^{4g} X_1\mathcal{H}_2 \wedge *_3 \mathcal{H}_2 + \tfrac{1}{2} \frac{e^{-4g}}{X_1} \left( a_1^2 + 4 X_1^3 e^{4g} (X_2 + X_3) \right) \mathcal{B}_1 \wedge *_3 \mathcal{B}_1.  
\ee
This term, when added to the original Lagrangian (\ref{Einsteinact}), allows one to support a null-warped AdS$_3$ vacuum in addition to the original AdS$_3$ vacuum. This new vacuum corresponds to the result of the TsT transformations discussed in section \ref{sec:TsT}, however since we have compromised the supersymmetry condition (\ref{cond_susy}) through the above identifications, supersymmetry will be broken. 

\subsection*{Supersymmetric vacuum} 
Taking into account our supersymmetry condition (\ref{cond_susy}), it is possible to find deformations that preserve supersymmetry. If one combines the TsT generated solution (\ref{gensol1}), (\ref{gensol2}) with S-duality, the $g_{++}$ term of the metric will be proportional to the following expression 
\be
\label{susy_g++}
g_{++} \propto \frac{\mu_1^2}{X_1} (\lambda_1^2 + \lambda_2^2)+ \frac{\mu_2^2}{X_2} (\lambda_3^2 + \lambda_4^2)+ \frac{\mu_3^2}{X_3} (\lambda_5^2 + \lambda_6^2),
\ee
where $\lambda_i$ are the now accustomed constants arising from TsT transformations. In particular, $\lambda_1, \lambda_3, \lambda_5$ are, up to a relabelling, the three constants appearing in the solution (\ref{gensol1}), whereas $\lambda_2, \lambda_4, \lambda_6$ are new constants that arise when one combines TsT with S-duality. To see how this happens, we can focus on the result of a single TsT with constant $\lambda_1$. The S-duality transformation interchanges the NS and RR two-form potentials and one notes that the resulting $B$-field has no $\varphi_1$ component, or in other words, $g=0$ in (\ref{Ansatz}). This means that applying TsT again along $\varphi_1$, this time with constant $\lambda_2$, we complete the first term in (\ref{susy_g++}). Repeating in similar fashion one fills out the remaining terms. 

Our supersymmetry analysis then tells us that we should impose the following constraints on the constants, 
\be
\label{susy_cond2}
0 = \lambda_1 +\lambda_3 + \lambda_5 =   \lambda_2 + \lambda_4 + \lambda_6. 
\ee
It is easier to first redefine the following 
\bea
 \left( \begin{array}{c} \lambda_{2i-1} \\ \lambda_{2i} \end{array} \right) &=&  \sqrt{\frac{X_i}{X_1} }\left( \begin{array}{cc} \cos \beta_i & \sin \beta_i \\ -\sin \beta_i & \cos \beta_i \end{array} \right) \left( \begin{array}{c} \lambda_1 \\ \lambda_2 \end{array} \right),
\eea
where have introduced two constant angles $\beta_i$, $i=2,3$, so that all dependence of $g_{++}$ on $\mu_i$ drops out. This ensures that $g_{++}$ is a constant and that a dimensionally reduced description can be found. One can then solve (\ref{susy_cond2}), thus ensuring a supersummetric solution, provided $\beta_2$ and $\beta_3$ are chosen so that 
\bea
\beta_2 = \sin^{-1} \left( - \sqrt{\frac{X_3}{X_1}} \sin \beta_3 \right), \quad \beta_3 = \frac{X_2 - X_1 - X_3}{2 \sqrt{X_1 X_3}}.  
\eea
Thus, to summarise, if we consider solutions generated via TsT and S-duality, we have 6 parameters with 4 constraints leading to a two-parameter family of supersymmetric null-warped AdS$_3$ solutions where the $g_{++}$ term is independent of the coordinates on the internal manifold.  

We now identify the corresponding contribution to the action, where we expect the relevant term to correspond to a massive vector. Inspired by our above analysis for the explicit supersymmetric solution, we make the identifications 
\bea
\dd b^i -2 B_1^i &=& \sqrt{\frac{X_1}{X_i}} \left[ \cos \beta_i (\dd \frak{b} - 2 \mathcal{B}_1) + \sin \beta_i (\dd \frak{c} -2 \mathcal{C}_1)  \right],  \nn
\dd c^i -2 C_1^i &=& \sqrt{\frac{X_1}{X_i}} \left[ -\sin \beta_i (\dd \frak{b} - 2 \mathcal{B}_1) + \cos \beta_i (\dd \frak{c} -2 \mathcal{C}_1)  \right], 
\eea
where now $i=1,2, 3$ and for consistency we require $\beta_1 = 0$. We employ similar identifications for the field strengths $H_2^i, G_2^i$ in terms of $\mathcal{H}_2 = \dd \mathcal{B}_1$ and $\mathcal{G}_2 = \dd \mathcal{C}_1$. The equations of motion can then be derived from the following Lagrangian 
\bea
\delta \mathcal{L}^{(3)} &=&  \tfrac{1}{8} {e^{-4g}}{ X_1^{-1}} \left( a_1^2 + 4 X_1^3 e^{4g} (X_2 + X_3) \right) \biggl[(\dd \frak{b}- 2 \mathcal{B}_1) \wedge *_3 (\dd \frak{b} -2 \mathcal{B}_1) \nn &+&  (\dd \frak{c} - 2 \mathcal{C}_1) \wedge *_3 (\dd \frak{c} - 2 \mathcal{C}_1) \biggr] + \tfrac{1}{2} e^{4g} X_1\left[ \mathcal{H}_2 \wedge *_3 \mathcal{H}_2 + \mathcal{G}_2 \wedge *_3 \mathcal{G}_2 \right]. 
\eea

In this expression we have retained the scalars, $\frak{b}, \frak{c}$, simply to illustrate that they are an expected component to any reduced theory. Further examples can be found in \cite{Karndumri:2013dca}.  Strictly speaking these scalars have been retained on the assumption that they are null, so they cannot depend on the radial coordinate of AdS$_3$ unless we give up on consistency.  However, we can consistently set them to a constant and truncate them out leaving us two massive vectors that can support supersymmetric null-warped AdS$_3$ solutions. 

Since neither of these produce Chern-Simons terms, which typically result from isometry gaugings in three dimensions, they do not contribute moment maps to the T tensor and, as a result, the T tensor is the same with our without the additional Lagrangian terms above. Thus, for the two-parameter class of supersymmetric null-warped AdS$_3$ solutions described by the above three-dimensional action, $c$-extremization picks out the correct R symmetry and central charge.

\section{Discussion} 
\label{sec:conclusions}
In this paper we have explored various aspects of supersymmetry for null-warped AdS$_3$ spacetimes. At the level of the superalgebra, we have exhibited subalgebras of the Lie superalgebra psu($1,1|2) \oplus$ psu (1,1$|$2), which reconcile Schr\"{o}dinger symmetry with supersymmetry. As expected, we have noted the existence of a maximal super Schr\"{o}dinger superalgebra with twelve supersymmetries, corresponding to a light-cone compactification of AdS$_3 \times$ S$^3$. Using projection conditions, we have identified an example with six supersymmetries and highlighted the existence of exotic superalgebras with less supersymmetry that are a direct consequence of the fact that we are working in lower dimensions. 

For the superalgebra with six supersymmetries, we have provided a string theory construction in terms of a deformation of an intersecting D3-brane solution to type IIB supergravity. We believe this is the first example of a null-warped AdS$_3$ solution with enhanced supersymmetry and hope it is a stepping stone to elucidating properties of the dual CFT. Though we are acutely aware that it is difficult to find solutions with less supersymmetry, it would be interesting to find an example of a geometry without kinematical supersymmetries, since its existence is hinted at through our superalgebra analysis. 

We have noted that TsT transformations acting on supersymmetric AdS$_3$ solutions to type IIB supergravity typically result in solutions where there are no superconformal Killing spinors and the number of Poincar\'e Killing spinors is halved.  It is interesting that these appear to be the same solutions one gets from twists of five-dimensional Schr\"{o}dinger solutions that exhibit supersymmetry enhancement. This raises a pertinent question about whether supersymmetry enhancement is actually possible for such classes of null-warped AdS$_3$ solutions.  One would need to find the analogue of the harmonic and primitive $(1,1)$-form that permits supersymmetry enhancement when the internal space is a Calabi-Yau cone. Alternatively, one could consider diagonal terms in the metric of the form $g_{+m}$, where $m$ denotes an internal direction \cite{Ooguri:2009cv, Jeong:2009aa}. 

In addition to the explicit example based on S$^5$ considered in \cite{Donos:2009zf}, we have identified two further type IIB Schr\"{o}dinger geometries, one based on $T^{1,1}$ in five dimensions, and the other on S$^3 \times$ CY$_2$ in three dimensions, where the Schr\"{o}dinger metric becomes independent of the internal geometry. This hints at the existence of a lower-dimensional theory, which has yet to be identified, that permits supersymmetry enhancement. Such a theory may serve as a setting to study classes of supersymmetric solutions with null-warped AdS$_3$ (Schr\"{o}dinger) near-horizons. With the dynamical exponent $z=4$, as we point out in appendix E, one can certainly find supersymmetric solutions corresponding to a large class of supertubes, so there may be some hope here. Separately, supersymmetric null-warped AdS$_3$ solutions have appeared in theories with Lorentz Chern-Simons terms \cite{Nilsson:2013fya,Deger:2013yla} and it has been observed in Ref. \cite{Deger:2013yla} that supersymmetry is not enhanced. We hope to address the identification of these lower-dimensional theories in future work to outline the minimal field content one requires for a geometric realisation of the super Schr\"{o}dinger algebra. From the five-dimensional perspective, it may also be worth investigating (now that we have found examples based on S$^5$ and $T^{1,1}$) if one can find Schr\"{o}dinger deformations for generic Sasaki-Einstein manifolds that allow a purely five-dimensional description, or are they simply the preserve of coset manifolds. If so, four-dimensional analogues in M-theory based on the Sasaki-Einstein manifolds S$^7$ and $Q^{1,1,1}$ (see \cite{Nilsson:1984bj}) are to be expected. 

A lower-dimensional description for these solutions should also present an insight into a potential solution generating mechanism. We recall that Schr\"{o}dinger solutions with enhanced supersymmetry have to be constructed case by case and it is not known how one generates them. As is common in gauged supergravities, the lower-dimensional picture may highlight a non-trivial transformation, potentially of Ehlers type, through which they can be generated. 

In the last part of this paper, we have observed that the $c$-extremization prescription presented in \cite{Karndumri:2013iqa} can also be applied to null-warped AdS$_3$ solutions that have been generated via TsT. It would be a considerable improvement on the treatment presented here if a construction with enhanced supersymmetry could be found, since in that case, we would be able to understand the R symmetry from the field theory perspective. In the light of the work of \cite{Detournay:2012pc} on warped CFTs with only SL($2, \mathbb{R}) \times$ U($1)$ isometry, by further incorporating supersymmetry, it may be hoped that  one can also find a field theory treatment of $c$-extremization for such a class of theories. It is also an obvious open direction to consider how $a$-maximization \cite{Intriligator:2003jj} may work in the non-relativistic setting. We hope to return to this question in future work. 

Finally, though we have not touched upon the subject here, now that we have an example of null-warped AdS$_3$ with enhanced supersymmetry, the identification of the dual field theory is a pressing concern. Dipole theories preserving supersymmetry are known  and the suggested gravity duals \cite{Bergman:2001rw} all involve deformations of the internal geometry. In contrast, here we witness no deformation of the internal geometry, so it is unlikely the field theory description corresponds to a dipole theory. Thus, the candidate dual field theory, if one exists, should be something new and it remains to be seen if a convincing candidate can be found. 

\section*{Acknowledgements} 
We have enjoyed conversations with K. Balasubramanian, F. Benini, D. Hofman, K. Jensen, P. Karndumri, D. S. Park, J. Schmude, E. Sezgin and H. Yavartanoo. E. \'O C would like to acknowledge the hospitality of the Simons Center for Geometry and Physics during the recent \textit{Aspects of Supergravity} workshop. J. J. is supported in part by the National Research Foundation of Korea (NRF) grant funded by the Korea government (MEST), with the grant number 2012046278 and the grant number 2013-110892. E. \'O C is supported by Marie Curie grant 328625 ``T-Dualities". 
\appendix 

\section{Super Schr\"odinger algebras}

In this Appendix, we consider super Schr\"odinger algebras as subalgebras of 
psu(1,1$|$2)$\oplus$ psu(1,1$|$2)\,. Here the superalgebra psu(1,1$|$2)$\oplus$ psu(1,1$|$2) 
describes the superconformal symmetry of the AdS$_3\times$S$^3$ geometry. 
We first give the algebraic relations of psu(1,1$|$2)$\oplus$ psu(1,1$|$2)\,. 
Then we show the super Schr\"odinger algebra preserving maximal supersymmetries. 
Finally subalgebras of the maximally supersymmetric Schr\"odinger algebra. 

\subsection*{The superalgebra psu(1,1$|$2)$\oplus$ psu(1,1$|$2)}

Let us introduce the algebraic relations of psu(1,1$|$2)$\oplus$ psu(1,1$|$2)\,. 

The bosonic part is given by 
\begin{eqnarray}
&& [D,P_{\mu}]=P_{\mu}\,, \quad [D,K_{\mu}]=-K_{\mu}\,, \quad [D,J_{\mu\nu}]=0\,,  \quad [P_{\mu},K_{\nu}] 
= 2 (\eta_{\mu\nu}\,D + J_{\mu\nu})\,, \nonumber \\ 
&& [P_{\rho}, J_{\mu\nu}] = \eta_{\mu\rho}\,P_{\nu} - \eta_{\nu\rho}\,P_{\mu}\,, \quad 
[K_{\rho},J_{\mu\nu}] = \eta_{\mu\rho}\,K_{\nu} - \eta_{\nu\rho}\,K_{\mu}\,, \nonumber \\ 
&& [R_{a},R_{b}] = -N_{ab}\,, \quad 
[R_{c},N_{ab}] = \delta_{ca}R_{b} - \delta_{cb}R_{a}\,,  \nonumber \\ 
&& [N_{cd},N_{ab}] = \delta_{ca}N_{db} + \delta_{db}N_{ca} 
- \delta_{cb}N_{da} - \delta_{da}N_{cb}\,. 
\end{eqnarray}
The other commutators vanish. 
Here the indices $\mu,\nu,\rho=0,1$ denote two-dimensional spacetime 
with the metric $\eta_{00}=-1$ and $\eta_{11}=1$\,. 
The indices $a,b,c,d = 1,2,3$ denote the internal symmetry su(2)$\oplus$su(2)\,. 
For later convenience, so(4) is decomposed into two su(2)'s.  

\medskip 

The mixed commutation relations of the bosonic generators and the supercharges 
are 
\begin{eqnarray}
&& [P_{\mu}, Q^I_{\alpha\hat{\alpha}}] = 0\,, \quad [K_{\mu}, S^I_{\alpha\hat{\alpha}}] = 0\,, \quad 
[D,Q^I_{\alpha\hat{\alpha}}] = \frac{1}{2}Q^I_{\alpha\hat{\alpha}}\,, \quad 
[D,S^I_{\alpha\hat{\alpha}}] = -\frac{1}{2}S^I_{\alpha\hat{\alpha}}\,, \nonumber \\ 
&& [J_{01},Q^I_{\alpha\hat{\alpha}}] = \frac{i}{2}(\hat{\gamma}_{01})_{\hat{\alpha}}^{~\hat{\beta}} Q^I_{\alpha\hat{\beta}}\,, 
\qquad  [J_{01},S^I_{\alpha\hat{\alpha}}] = \frac{i}{2}(\hat{\gamma}_{01})_{\hat{\alpha}}^{~\hat{\beta}} S^I_{\alpha\hat{\beta}}\,, 
\nonumber \\
&& [K_{\mu},Q^I_{\alpha\hat{\alpha}}] = i(\hat{\gamma}_{\mu})_{\hat{\alpha}}^{~\hat{\beta}}S^I_{\alpha\hat{\beta}}\,, \qquad 
[K_{\mu},S^I_{\alpha\hat{\alpha}}] = i(\hat{\gamma}_{\mu})_{\hat{\alpha}}^{~\hat{\beta}}Q^I_{\alpha\hat{\beta}}\,, \nonumber \\ 
&& [R_a, S^I_{\alpha\hat{\alpha}}] = \frac{i}{2}(i\hat{\gamma}_{01})_{\hat{\alpha}}^{~\hat{\beta}}(\gamma_{a})_{\alpha}^{~\beta}S^I_{\beta\hat{\beta}}\,, 
\qquad  [R_a, Q^I_{\alpha\hat{\alpha}}] = -\frac{i}{2}(i\hat{\gamma}_{01})_{\hat{\alpha}}^{~\hat{\beta}}(\gamma_{a})_{\alpha}^{~\beta}Q^I_{\beta\hat{\beta}}\,, 
\nonumber \\ 
&& [N_{ab}, S^I_{\alpha\hat{\alpha}}] = \frac{i}{2}\hat{C}_{\hat{\alpha}}^{~\hat{\beta}}(\gamma_{ab})_{\alpha}^{~\beta}S^I_{\beta\hat{\beta}}\,, \qquad 
[N_{ab}, Q^I_{\alpha\hat{\alpha}}] = \frac{i}{2}\hat{C}_{\hat{\alpha}}^{~\hat{\beta}}(\gamma_{ab})_{\alpha}^{~\beta}Q^I_{\beta\hat{\beta}}\,. 
\label{BQ}
\end{eqnarray}
Here $I=1,2$\,, $\hat{\alpha}=1,2$\,, $\alpha=1,2$ and the charge conjugation matrices are defined as 
\[
\hat{C}_{\hat{\alpha}\hat{\beta}} = \epsilon_{\hat{\alpha}\hat{\beta}}\,, \qquad C_{\alpha\beta} = \epsilon_{\alpha\beta}\,. 
\]
The gamma matrices are given by 
\begin{eqnarray}
&& (\hat{\gamma}_0)_{\hat{\alpha}}^{~\hat{\beta}} = i\sigma_2\,, \qquad 
(\hat{\gamma}_1)_{\hat{\alpha}}^{~\hat{\beta}}  = -\sigma_1\,, \qquad 
(\hat{\gamma}_{01})_{\hat{\alpha}}^{~\hat{\beta}}  \equiv 
\frac{i}{2}[\hat{\gamma}_0,\hat{\gamma}_1] 
= -i\sigma_3\,, \\
&& (\gamma_1)_{\alpha}^{~\beta}  = \sigma_1\,, \qquad 
(\gamma_2)_{\alpha}^{~\beta} = \sigma_3\,, \qquad 
(\gamma_3)_{\alpha}^{~\beta} = \sigma_2\,, 
\end{eqnarray}
where $\sigma_i~(i=1,2,3,)$ are the standard Pauli matrices. 
The spinor convention is\footnote{Here we follow the convention and notation utilized in \cite{ADO}.} 
\[
\psi_{\alpha} = \psi^{\beta}\epsilon_{\beta\alpha}\,, \qquad \psi^{\alpha} = \epsilon^{\alpha\beta}\psi_{\beta}\,. 
\]

\medskip 

Finally, the commutation relations including only the supercharges are given by 
\begin{eqnarray}
&& \{Q^I_{\alpha\hat{\alpha}}, Q^J_{\beta\hat{\beta}}\} = \frac{i}{2}\epsilon^{IJ}C_{\alpha\beta}
(\hat{\gamma}^{\mu}\hat{C})_{\hat{\alpha}\hat{\beta}}P_{\mu}\,, \qquad 
 \{S^I_{\alpha\hat{\alpha}}, S^J_{\beta\hat{\beta}}\} = \frac{i}{2}\epsilon^{IJ}C_{\alpha\beta}
(\hat{\gamma}^{\mu}\hat{C})_{\hat{\alpha}\hat{\beta}}K_{\mu}\,, \nonumber \\ 
&& \{S^I_{\alpha\hat{\alpha}},Q^J_{\beta\hat{\beta}}\} = \frac{i}{2}\epsilon^{IJ}
\biggl(
C_{\alpha\beta} \left[ i\hat{C}_{\hat{\alpha}\hat{\beta}}D + (\hat{C}\hat{\gamma}_{01})_{\hat{\alpha}\hat{\beta}}J_{01}
\right]  \nonumber \\ 
&& \hspace*{3cm}
+ \frac{1}{2}\hat{C}_{\hat{\alpha}\hat{\beta}}N_{ab}(\gamma^{ab}C)_{\alpha\beta} -i(\hat{\gamma}_{01}\hat{C})_{\hat{\alpha}\hat{\beta}}
R_a(\gamma^aC)_{\alpha\beta}\biggr)\,. \label{QQ}
\end{eqnarray}

In the next subsection, we will consider super Schr\"odinger algebras by using the algebraic relations 
introduced here.

\subsection*{Super Schr\"odinger algebras}

Let us consider super Schr\"odinger algebras. The first goal is to find out the maximally supersymmetric 
Schr\"odinger algebra by following the strategy argued in \cite{SY1,SY2}\footnote{
In \cite{SY1,SY2}, super Schr\"odinger algebras are considered from psu($2,2|4$), 
osp(8$|$4) and osp($8^{\ast}|4$). For generalizations to the $z\neq 2$ case and 
classifications of super Lifshitz algebras, see \cite{SY3}. }. 
Then it enables us to consider less supersymmetric Schr\"odinger algebras 
as subalgebras of the maximal one. 

\subsubsection*{The bosonic part}

We first consider the bosonic part of super Schr\"odinger algebras. For this purpose, 
it is necessary to introduce the light-cone coordinate. 
Our light-cone convention is the following: 
\begin{eqnarray}
P_{\pm} = \frac{1}{\sqrt{2}}(P_1 \pm P_0)\,, \qquad  K_{\pm} = \frac{1}{\sqrt{2}}(K_1 \pm K_0)\,. 
\end{eqnarray}
The dilatation generator with $z=2$ is defined as 
\begin{eqnarray}
\tilde{D} \equiv D + J_{01}\,. 
\end{eqnarray}
It is convenient to use the notation:
\[
 H \equiv P_+\,, \qquad M \equiv P_-\,, \qquad C \equiv K_-/2\,.
\]
Then the bosonic Schr\"odinger algebra is obtained as a subalgebra,  
\begin{eqnarray}
[\tilde{D},H]=2H\,, \quad [\tilde{D},C]=-2C\,, \quad [H,C]=-\tilde{D}\,.
\end{eqnarray}
Here $M$ is a center. 
It is easy to check the Jacobi identity. 
In the present case there is no spatial translation and rotation, Galilean boost. 
The resulting algebra is nothing but $SL(2,\mathbb{R})\times U(1)$\,. 

\subsubsection*{The fermionic part} 

The next is to argue the fermionic part. The bosonic Schr\"odinger algebra is realized as a subalgebra 
of the conformal algebra. Hence, according to the restriction, it is necessary to project out some of 
the supersymmetries contained in psu(1,1$|$2)$\oplus$ psu(1,1$|$2)\,.  

Let us first see the anti-commutators in (\ref{QQ}). 
The first anti-commutator is obviously irrelevant. 
The second anti-commutator gives a constraint. It can be rewritten as 
\begin{eqnarray}
\{S^I_{\alpha\hat{\alpha}}, S^J_{\beta\hat{\beta}}\} = \frac{i}{2\sqrt{2}}\epsilon^{IJ}C_{\alpha\beta}
\left(\left[
K_+ (\hat{\gamma}_1-\hat{\gamma}_0) + K_-(\hat{\gamma}_1 + \hat{\gamma}_0)
\right]
\hat{C}
\right)_{\hat{\alpha}\hat{\beta}}\,. 
\end{eqnarray}
Note that the unwanted generator $K_+$\,, so as to close the Schr\"odinger algebra, 
is multiplied by the matrix 
\[
\left[(\hat{\gamma}_1-\gamma_0)\hat{C}\right]_{\hat{\alpha}\hat{\beta}} = 
\begin{pmatrix} 2 & 0 \\ 0 & 0
\end{pmatrix}\,.  
\]
In order to remove $K_+$\,, the component $S^I_{\alpha 1}$ has to be eliminated. 
It is useful to decompose $S^I_{\alpha\hat{\alpha}}$ with the projection operator, 
\begin{eqnarray}
S^I_{\alpha\hat{\alpha}} &\equiv& S^{(-),I}_{\alpha\hat{\alpha}} + S^{(+),I}_{\alpha\hat{\alpha}} \nonumber \\ 
&=& \left(\mathcal{\widehat{P}}_-\right)_{\hat{\alpha}}^{~\hat{\beta}} S^I_{\alpha\hat{\beta}} + 
 \left(\mathcal{\widehat{P}}_+\right)_{\hat{\alpha}}^{~\hat{\beta}} S^I_{\alpha\hat{\beta}}\,,  
\end{eqnarray}
where we have defined the projection operators, 
\[
(\mathcal{\widehat{P}}_{\pm})_{\hat{\alpha}}^{~\hat{\beta}} \equiv 
\frac{1}{2} \left(1 \pm i\hat{\gamma}_{01}\right)_{\hat{\alpha}}^{~\hat{\beta}}\,. 
\]
Then, by using the projected supercharge 
\[
S^{(-),I}_{\alpha\hat{\alpha}} = \left(\mathcal{\widehat{P}}_-\right)_{\hat{\alpha}}^{~\hat{\beta}} S^I_{\alpha\hat{\beta}}\,,
\]
the super Schr\"odinger algebra is closed, and the anti-commutator is rewritten as 
\begin{eqnarray}
\{S^{(-),I}_{\alpha\hat{\alpha}}\,, S^{(-),J}_{\beta\hat{\beta}} \} = i\sqrt{2}\,\epsilon^{IJ}C_{\alpha\beta}\, C 
\left[\hat{\gamma}_0\widehat{\mathcal{P}}_+\hat{C}
\right]_{\hat{\alpha}\hat{\beta}}\,. 
\end{eqnarray}

The next is the last anti-commutator. Noting that 
\begin{eqnarray}
i\hat{C}_{\hat{\alpha}\hat{\beta}}D + (\hat{C}\hat{\gamma}_{01})_{\hat{\alpha}\hat{\beta}}J_{01} = 
\begin{pmatrix}
0 & i(D-J_{01}) \\ 
-i(D+J_{01}) && 0 
\end{pmatrix}
\,,
\end{eqnarray}
the last anti-commutator is rewritten as 
\begin{eqnarray}
\{S^{(-),I}_{\alpha\hat{\alpha}}\,, Q^J_{\beta\hat{\beta}}\} 
&=& \frac{i}{2}\epsilon^{IJ}\biggl[
-C_{\alpha\beta} \left(\mathcal{\widehat{P}}_-\hat{C}\right)_{\hat{\alpha}\hat{\beta}} i \tilde{D} \nonumber \\ 
&& \hspace*{1.5cm} + \frac{1}{2}(\mathcal{\widehat{P}}_-\hat{C})_{\hat{\alpha}\hat{\beta}}N_{ab}(\gamma^{ab}C)_{\alpha\beta} 
+ (\mathcal{\widehat{P}}_-\hat{C})_{\hat{\alpha}\hat{\beta}}
R_a(\gamma^aC)_{\alpha\beta}\biggr]\,.
\end{eqnarray}

\medskip 

The remaining task is to check the commutator including the bosonic and fermionic generators. 
It is easy to show the following relations, 
\begin{eqnarray}
&& [D,S^{(-),I}_{\alpha\hat{\alpha}}] = -\frac{1}{2}S^{(-),I}_{\alpha\hat{\alpha}}\,, \qquad 
[J_{01},S^{(-),I}_{\alpha\hat{\alpha}}] = \frac{i}{2}(\hat{\gamma}_{01})_{\hat{\alpha}}^{~\hat{\beta}}
S^{(-),I}_{\alpha\hat{\beta}}\,, \nonumber \\ 
&& [C,Q^I_{\alpha\hat{\alpha}}] = -\frac{i}{\sqrt{2}}(\hat{\gamma}_0)_{\hat{\alpha}}^{~\hat{\beta}}
S^{(-),I}_{\alpha\hat{\beta}}\,, \qquad 
[C,S^{(-),I}_{\alpha\hat{\alpha}}] =  0\,, 
\nonumber \\ 
&& [R_a,S^{(-),I}_{\alpha\hat{\alpha}}] = - \frac{i}{2}
(\gamma_{a})_{\alpha}^{~\beta}S^{(-),I}_{\beta\hat{\alpha}}\,, \qquad 
[R_a, Q^{I}_{\alpha\hat{\alpha}}] = -\frac{i}{2}(i\hat{\gamma}_{01})_{\hat{\alpha}}^{~\hat{\beta}}(\gamma_{a})_{\alpha}^{~\beta}
Q^I_{\beta\hat{\beta}}\,, \nonumber \\ 
&& [N_{ab},S^{(-),I}_{\alpha\hat{\alpha}}] = \frac{i}{2}\hat{C}_{\hat{\alpha}}^{~\hat{\beta}}(\gamma_{ab})_{\alpha}^{~\beta}S^{(-),I}_{\beta\hat{\beta}}\,.
\nonumber 
\end{eqnarray}
Thus the Schr\"odinger algebra is closed with the supercharges $Q^I_{\alpha\hat{\alpha}}$ (8 Poincare SUSY) 
and $S^{(-),I}_{\alpha\hat{\alpha}}$ (4 conformal SUSY)\,. That is, in total, 12=(16 $\times$ 3/4) supersymmetries are preserved. 

\subsubsection*{The maximal super Schr\"odinger algebra}

It is valuable to summarize the (anti-)commutation relations of the maximally supersymmetric 
Schr\"odinger algebra. 

The commutation relations of the maximal super Schr\"odinger algebra: 
\begin{eqnarray}
&& [\tilde{D},H]=2H\,, \quad [\tilde{D},C]=-2C\,, \quad [H,C]=-\tilde{D}\,, \nonumber \\ 
&& 
[\tilde{D},Q^I_{\alpha\hat{\alpha}}] = (\mathcal{\widehat{P}}_+)_{\hat{\alpha}}^{~\hat{\beta}}Q^I_{\alpha\hat{\beta}}\,, 
\quad [\tilde{D},S^{(-),I}_{\alpha\hat{\alpha}}] = 
-S^{(-),I}_{\alpha\hat{\beta}}\,, \nonumber \\ 
&& [C,Q^I_{\alpha\hat{\alpha}}] = -\frac{i}{\sqrt{2}}(\hat{\gamma}_0)_{\hat{\alpha}}^{~\hat{\beta}}
S^{(-),I}_{\alpha\hat{\beta}}\,, \quad 
[C,S^{(-),I}_{\alpha\hat{\alpha}}] =  0\,,
\nonumber \\ 
&& [R_a,S^{(-),I}_{\alpha\hat{\alpha}}] =  
-\frac{i}{2}
(\gamma_{a})_{\alpha}^{~\beta}S^{(-),I}_{\beta\hat{\alpha}}\,, \quad 
[R_a, Q^{I}_{\alpha\hat{\alpha}}] = -\frac{i}{2}(i\hat{\gamma}_{01})_{\hat{\alpha}}^{~\hat{\beta}}(\gamma_{a})_{\alpha}^{~\beta}
Q^I_{\beta\hat{\beta}}\,, \nonumber \\ 
&& [N_{ab},S^{(-),I}_{\alpha\hat{\alpha}}] = \frac{i}{2}\hat{C}_{\hat{\alpha}}^{~\hat{\beta}}(\gamma_{ab})_{\alpha}^{~\beta}S^{(-),I}_{\beta\hat{\beta}}\,, 
\quad [N_{ab}, Q^I_{\alpha\hat{\alpha}}] = \frac{i}{2}\hat{C}_{\hat{\alpha}}^{~\hat{\beta}}(\gamma_{ab})_{\alpha}^{~\beta}Q^I_{\beta\hat{\beta}}\,, 
\nonumber \\ 
&& \{Q^I_{\alpha\hat{\alpha}}, Q^J_{\beta\hat{\beta}}\} = \frac{i}{2}\epsilon^{IJ}C_{\alpha\beta} \left[
- \sqrt{2} H (\hat{\gamma}_0\widehat{\mathcal{P}}_-\hat{C})_{\hat{\alpha}\hat{\beta}} 
+ \sqrt{2} M (\hat{\gamma}_0\widehat{\mathcal{P}}_+\hat{C})_{\hat{\alpha}\hat{\beta}}\right]\,, 
 \nonumber \\
&& \{S^{(-),I}_{\alpha\hat{\alpha}}\,, S^{(-),J}_{\beta\hat{\beta}} \} = i\sqrt{2}\,\epsilon^{IJ}C_{\alpha\beta}\, C 
\left[\hat{\gamma}_0\widehat{\mathcal{P}}_+\hat{C}
\right]_{\hat{\alpha}\hat{\beta}}\,, \nonumber \\ 
&& \{S^{(-),I}_{\alpha\hat{\alpha}}\,, Q^J_{\beta\hat{\beta}}\} 
= \frac{i}{2}\epsilon^{IJ}\biggl[
-C_{\alpha\beta} \left(\mathcal{\widehat{P}}_-\hat{C}\right)_{\hat{\alpha}\hat{\beta}} i \tilde{D} \nonumber \\ 
&& \hspace*{4.5cm} + \frac{1}{2}(\mathcal{\widehat{P}}_-\hat{C})_{\hat{\alpha}\hat{\beta}}N_{ab}(\gamma^{ab}C)_{\alpha\beta} 
+ (\mathcal{\widehat{P}}_-\hat{C})_{\hat{\alpha}\hat{\beta}}
R_a(\gamma^aC)_{\alpha\beta}\biggr]\,.
\end{eqnarray}
Here we have used the formula,
\begin{eqnarray}
(\hat{\gamma}^{\mu}\hat{C})_{\hat{\alpha}\hat{\beta}} P_{\mu} = 
- \sqrt{2} H (\hat{\gamma}_0\widehat{\mathcal{P}}_-\hat{C})_{\hat{\alpha}\hat{\beta}} 
+ \sqrt{2} M (\hat{\gamma}_0\widehat{\mathcal{P}}_+\hat{C})_{\hat{\alpha}\hat{\beta}}\,.  
\end{eqnarray}
The remaining SUSY is 8 Poincare SUSY and 4 conformal SUSY. 

\subsection*{Super subalgebras}

Then we consider an example of less supersymmetric Schr\"odinger algebras, 
which preserves 4 Poincare SUSY and 2 conformal SUSY. 
Then the internal symmetry is taken $N_{13}$ and $R_2$ as the non-vanishing component. 
The other components of $N_{ab}$ and $R_a$ are zero. According to this choice, 
the original $SU(2)_{\rm L} \times SU(2)_{\rm R}$ symmetry is broken to 
$U(1)_{\rm L} \times U(1)_{\rm R}$\,. 

\medskip

The additional projection condition is given with the projection operators
\[
(\mathcal{P}_{\pm})_{\alpha}^{~\beta} \equiv \frac{1}{2}\left(1 \pm \gamma_2 \right)_{\alpha}^{~\beta} 
\]
as follows: 
For conformal SUSY, 
\begin{eqnarray}
S^{(-),I}_{\alpha\hat{\alpha}} &=& S^{[+](-),I}_{\alpha\hat{\alpha}} + S^{[-](-),I}_{\alpha\hat{\alpha}}  \nonumber \\ 
&=& (\mathcal{P}_+)_{\alpha}^{~\beta} S^{(-),I}_{\beta\hat{\alpha}} 
+ (\mathcal{P}_-)_{\alpha}^{~\beta} S^{(-),I}_{\beta\hat{\alpha}}\,.  
\end{eqnarray}
For Poincare SUSY, 
\begin{eqnarray}
Q^{I}_{\alpha\hat{\alpha}} &=& Q^{[+],I}_{\alpha\hat{\alpha}} + Q^{[-],I}_{\alpha\hat{\alpha}}  \nonumber \\ 
&=& (\mathcal{P}_+)_{\alpha}^{~\beta} Q^{I}_{\beta\hat{\alpha}} 
+ (\mathcal{P}_-)_{\alpha}^{~\beta} Q^{I}_{\beta\hat{\alpha}}\,.  
\end{eqnarray}
Here, to make the anti-commutator $\{Q,Q\}$ non-vanishing, 
let us keep the components
\[
Q^{[+],1}_{\alpha\hat{\alpha}}\,, \quad Q^{[-],2}_{\beta\hat{\beta}}\,, \quad 
S^{[+](-),1}_{\alpha\hat{\alpha}}\,, \quad S^{[-](-),2}_{\beta\hat{\beta}}\,. 
\]

The commutation relations of the resulting super Schr\"odinger subalgebra: 
\begin{eqnarray}
&& [\tilde{D},H]=2H\,, \quad [\tilde{D},C]=-2C\,, \quad [H,C]=-\tilde{D}\,, \nonumber \\ 
&& 
[\tilde{D},Q^{[+],1}_{\alpha\hat{\alpha}}] = (\mathcal{\widehat{P}}_+)_{\hat{\alpha}}^{~\hat{\beta}}Q^{[+],1}_{\alpha\hat{\beta}}\,, 
\quad [\tilde{D},Q^{[-],2}_{\alpha\hat{\alpha}}] = (\mathcal{\widehat{P}}_+)_{\hat{\alpha}}^{~\hat{\beta}}Q^{[-],2}_{\alpha\hat{\beta}}\,, 
\nonumber \\ 
&& [\tilde{D},S^{[+](-),1}_{\alpha\hat{\alpha}}] = 
-S^{[+](-),1}_{\alpha\hat{\beta}}\,, \quad 
[\tilde{D},S^{[-](-),2}_{\alpha\hat{\alpha}}] = -S^{[-](-),2}_{\alpha\hat{\beta}}\,, 
\nonumber \\ 
&& [C,Q^{[+],1}_{\alpha\hat{\alpha}}] = -\frac{i}{\sqrt{2}}(\hat{\gamma}_0)_{\hat{\alpha}}^{~\hat{\beta}}
S^{[+](-),1}_{\alpha\hat{\beta}}\,, \quad 
[C,Q^{[-],2}_{\alpha\hat{\alpha}}] = -\frac{i}{\sqrt{2}}(\hat{\gamma}_0)_{\hat{\alpha}}^{~\hat{\beta}}
S^{[-](-),2}_{\alpha\hat{\beta}}\,, 
\nonumber \\ 
&& [R_2,S^{[+](-),1}_{\alpha\hat{\alpha}}] =  
-\frac{i}{2}(\gamma_{2})_{\alpha}^{~\beta}S^{[+](-),1}_{\beta\hat{\alpha}}\,, \quad 
[R_2,S^{[-](-),2}_{\alpha\hat{\alpha}}] =  
-\frac{i}{2}(\gamma_{2})_{\alpha}^{~\beta}S^{[-](-),2}_{\beta\hat{\alpha}}\,, \nonumber \\
&& 
[R_2, Q^{[+],1}_{\alpha\hat{\alpha}}] = -\frac{i}{2}(i\hat{\gamma}_{01})_{\hat{\alpha}}^{~\hat{\beta}}(\gamma_{2})_{\alpha}^{~\beta}
Q^{[+],1}_{\beta\hat{\beta}}\,, \quad 
[R_2, Q^{[-],2}_{\alpha\hat{\alpha}}] = -\frac{i}{2}(i\hat{\gamma}_{01})_{\hat{\alpha}}^{~\hat{\beta}}(\gamma_{2})_{\alpha}^{~\beta}
Q^{[-],2}_{\beta\hat{\beta}}\,, 
\nonumber \\ 
&& [N_{13},S^{[+](-),1}_{\alpha\hat{\alpha}}] = \frac{i}{2}\hat{C}_{\hat{\alpha}}^{~\hat{\beta}}(\gamma_{13})_{\alpha}^{~\beta}
S^{[+](-),1}_{\beta\hat{\beta}}\,, 
\quad [N_{13},S^{[-](-),2}_{\alpha\hat{\alpha}}] = \frac{i}{2}\hat{C}_{\hat{\alpha}}^{~\hat{\beta}}(\gamma_{13})_{\alpha}^{~\beta}
S^{[-](-),2}_{\beta\hat{\beta}}\,, 
\nonumber \\
&& [N_{13}, Q^{[+],1}_{\alpha\hat{\alpha}}] = \frac{i}{2}\hat{C}_{\hat{\alpha}}^{~\hat{\beta}}(\gamma_{13})_{\alpha}^{~\beta}
Q^{[+],1}_{\beta\hat{\beta}}\,, \quad 
[N_{13}, Q^{[-],2}_{\alpha\hat{\alpha}}] = \frac{i}{2}\hat{C}_{\hat{\alpha}}^{~\hat{\beta}}(\gamma_{13})_{\alpha}^{~\beta}
Q^{[-],2}_{\beta\hat{\beta}}\,, 
\nonumber \\ 
&& \{Q^{[+],1}_{\alpha\hat{\alpha}}, Q^{[-],2}_{\beta\hat{\beta}}\} = \frac{i}{2}(\mathcal{P}_+C)_{\alpha\beta} \left[
- \sqrt{2} H (\hat{\gamma}_0\widehat{\mathcal{P}}_-\hat{C})_{\hat{\alpha}\hat{\beta}} 
+ \sqrt{2} M (\hat{\gamma}_0\widehat{\mathcal{P}}_+\hat{C})_{\hat{\alpha}\hat{\beta}}\right]\,, 
 \nonumber \\
&& \{S^{[+](-),1}_{\alpha\hat{\alpha}}\,, S^{[-](-),2}_{\beta\hat{\beta}} \} = i\sqrt{2}\,(\mathcal{P}_+C)_{\alpha\beta}\, C 
\left[\hat{\gamma}_0\widehat{\mathcal{P}}_+\hat{C}
\right]_{\hat{\alpha}\hat{\beta}}\,, \nonumber \\ 
&& \{S^{[+](-),1}_{\alpha\hat{\alpha}}\,, Q^{[-],2}_{\beta\hat{\beta}}\} 
= \frac{i}{2}\biggl[
-(\mathcal{P}_+C)_{\alpha\beta} \left(\mathcal{\widehat{P}}_-\hat{C}\right)_{\hat{\alpha}\hat{\beta}} i \tilde{D} \nonumber \\ 
&& \hspace*{4.5cm} + \frac{1}{2}(\mathcal{\widehat{P}}_-\hat{C})_{\hat{\alpha}\hat{\beta}}N_{13}(\mathcal{P}_+C)_{\alpha\beta} 
+ (\mathcal{\widehat{P}}_-\hat{C})_{\hat{\alpha}\hat{\beta}}
R_2(\mathcal{P}_+C)_{\alpha\beta}\biggr]\,, \nonumber \\ 
&& \{S^{[-](-),2}_{\alpha\hat{\alpha}}\,, Q^{[+],1}_{\beta\hat{\beta}}\} 
= \frac{i}{2}\biggl[
(\mathcal{P}_-C)_{\alpha\beta} \left(\mathcal{\widehat{P}}_-\hat{C}\right)_{\hat{\alpha}\hat{\beta}} i \tilde{D} \nonumber \\ 
&& \hspace*{4.5cm} + \frac{1}{2}(\mathcal{\widehat{P}}_-\hat{C})_{\hat{\alpha}\hat{\beta}}N_{13}(\mathcal{P}_-C)_{\alpha\beta} 
+ (\mathcal{\widehat{P}}_-\hat{C})_{\hat{\alpha}\hat{\beta}}
R_2(\mathcal{P}_-C)_{\alpha\beta}\biggr]\,.
\end{eqnarray}

Similarly, it is possible to find out less supersymmetric Schr\"odinger algebras. 
For example, by keeping $Q_{\alpha\hat{\alpha}}^{[+](+),1}$,  $Q_{\beta\hat{\beta}}^{[-](+),2}$, 
$S^{[+](-),1}_{\alpha\hat{\alpha}}$ and $S^{[-](-),2}_{\beta\hat{\beta}}$\,, the resulting algebra 
preserves 2 dynamical supersymmetries and 2 conformal supersymmetries. 
The algebra of this type has not been found from psu($2,2|4$), 
osp(8$|$4) and osp($8^{\ast}|4$). The existence of such an algebra is based on 
the dimensionality of the present case we consider.  

\section{Killing spinor analysis}
We follow the supersymmetry conventions of \cite{Donos:2009zf}, which entails solving a gravitino and dilatino variation of the form 
\bea
\label{KSE1}
D_{M} \epsilon + \tfrac{i}{16} \slashed{F}_5 \Gamma_M \epsilon + \tfrac{1}{16} (\G_M \slashed{G}_3 + 2 \slashed{G}_3 \G_M) \epsilon^* &=& 0, \nn
\slashed{G}_3 \epsilon &=& 0. 
\eea
Here we work with a basis where the gamma matrices are real and $\epsilon^c = \epsilon^*$. We use the conventions $\G_{11} \epsilon = \G_{+-23456789} \epsilon = - \epsilon$. 

From the offset we have assumed that $f$ and $W$ are independent of the transverse $CY_2$. This means that $CY_2$ plays largely no role and the gravitino variation along these directions is satisfied provided 
\be
\label{en_susy_eq1}
D_{m} \epsilon = 0, \quad \G_{6789} \epsilon = - \epsilon, ~~\slashed{G}_3 \epsilon^*=0, 
\ee
where the $m=6, 7, 8, 9$ labels the $CY_2$ directions. 

The Killing spinor equation (KSE) along $M=-$ takes the form 
\be
\partial_{-} \epsilon =  \tfrac{1}{2} r (1- \gamma_{D3}) \G^r \G_{-} \epsilon, 
\ee
where we have defined $\gamma_{D3} = i \G^{+-67}$ \footnote{The Poincar\'e supersymmetries of the undeformed geometry, $f = W = 0$, correspond to the Killing spinors satisfying the projection condition $\gamma_{D3} \epsilon = \epsilon$. Here we use the subscript $D3$ to simply denote the fact that the AdS$_3$ geometries arise from wrapped D3-branes. In particular, when $CY_2 = T^4$, the AdS$_3$ solution is the near-horizon of two intersecting D3-branes.}. This equation implies $\partial_{-}^2 \epsilon = 0$, so that the Killing spinor is linear in $x^-$. This leads to the unique solution 
\be
\epsilon = \epsilon_0 + \tfrac{1}{2} r \G^r (x^- \G_-) (1+ \gamma_{D3}) \epsilon_0, 
\ee
where $\epsilon_0$ is independent of $x^-$. Following \cite{Donos:2009zf}, we can now decompose $\e_0$
\be
\e_0 = r^{-\frac{1}{2}} \G^r \e_+ + r^{\frac{1}{2}} \e_-, 
\ee
where $\g_{D3} \e_{\pm} = \pm \e_{\pm}$. The Killing spinor may then be rewritten as 
\be
\label{KS1}
\epsilon = r^{\frac{1}{2}} \e_- + \left[ r^{-\frac{1}{2}} \G^r - r^{\frac{1}{2}} (x^- \G_-) \right] \e_+.
\ee

This is essentially (2.13) of \cite{Donos:2009zf} when one notes that there are no additional spatial directions, so that the last term vanishes. 

From the dilatino variation and (\ref{en_susy_eq1}), we can now infer the following 
\bea
\label{proj_cond}
\G^+ \slashed{W} \e_{-} = \G^+ \slashed{W} \epsilon_-^*=  \G^{+} \slashed{W} \G^r \e_+ = \G^{+} \slashed{W} \G^r \e_+^* = 0. 
\eea

Moving onto $M= a$, $a=2, 3, 4, 5$, inserting our expression for the Killing spinor (\ref{KS1}) and decomposing under $\gamma_{D3}$, we get the following equations: 
\bea
\label{psi_R4_eq1} \nabla^4_a \e_+ + \tfrac{1}{8} \G^r \G^+ \slashed{W}( \G_{\alpha} e^{\alpha}_{a})  \G^r \e_+^{*} &=& 0, \\
\label{psi_R4_eq2} \nabla^4_a \e_- + \tfrac{1}{8} \G^+ \slashed{W}( \G_{\alpha} e^{\alpha}_{a})  \e_-^{*} &=& 0
\eea
where the superscript identifies the above covariant derivative as that of $\mathbb{R}^4$ with orthonormal frame 
\be
e^r = \dd r, ~~ e^{\alpha} = r \, \bar{e}^{\alpha},
\ee
where $\alpha = 3, 4, 5$ labels the orthonormal frame for $S^3$. These equations can again be mapped to (2.18) and (2.19) of \cite{Donos:2009zf} when one takes (\ref{proj_cond}) into account. The absence of spatial directions for our Schr\"{o}dinger solution mean that there is no analogue of (2.17) of \cite{Donos:2009zf}.  

The $M=+$ component of the KSE is 
\bea
\partial_{+} \epsilon + \tfrac{1}{4} r f \G^{+r} (1+ \gamma_{D3}) \e + \tfrac{1}{2} r \G^{-r} (1+ \gamma_{D3}) \e + \tfrac{1}{4} r^2 \slashed{W} \e^*  + \tfrac{1}{4} r^2 \G^+ \slashed{\partial} f \e = 0. 
\eea
Decomposing this further, one gets 
\bea
\label{psi_plus_eq1} \partial_+ \e_+ &=& 0, \\
\label{psi_plus_eq2} \partial_+ \e_- + \G^- \e_+ + \tfrac{1}{4} r \slashed{W} \G^r \e_+^* + \G^+ \left( \tfrac{1}{4} r \slashed{\partial} f \G^r + \tfrac{1}{2} f \right) \e_+ &=& 0,  \\
\label{psi_plus_eq3} \G^+ \slashed{\partial} f \e_- + \slashed{W} \e_-^* &=& 0, \\
\label{psi_plus_eq4} \G^+ \slashed{W} \e_+^* &=&0. 
\eea
Three of these equations are identical to (2.29)-(2.32) of \cite{Donos:2009zf}, up to the notable absence of an expression used to derive the result 
\be
\label{sec3+}
\G^+ \e_+ =0. 
\ee

This appears to be a marked difference with the same calculation in five dimensions. Here, (\ref{sec3+}) does not appear to follow obviously from a constraint derived from the KSE. This appears to allow room for the presence of exotic supersymmetry enhancement that does not fit into the usual pattern of kinematical, dynamical and superconformal supersymmetries. 

However, once this condition is imposed, the Killing spinors can be solved in the analogous fashion to \cite{Donos:2009zf}.  From (\ref{psi_R4_eq1}), we see that $\e_+$ is covariantly constant on $\mathbb{R}^4$
\be
\nabla^4_{a} \e_+ = 0. 
\ee

We can solve (\ref{psi_plus_eq2}) as 
\be
\e_- = \psi_- - x^+ \left( \G^- \eta_+ + \tfrac{1}{4} r \slashed{W} \G^r \eta_+^* \right), 
\ee
where $\psi_-$ is independent of $x^+$ and we have relabelled $\e_+ = \eta_+$. This expression can be shown to be compatible with (\ref{psi_R4_eq2}) provided\footnote{One can use (2.22) of \cite{Donos:2009zf} with $z=2$.}
\be
\label{psi_plus}\nabla^4_{a} \psi_- + \tfrac{1}{8} \G^+ \slashed{W} \G_a \psi_-^* = 0. 
\ee

We can now return to (\ref{psi_plus_eq3}), which implies 
\bea
\left(\slashed{\partial} f - \tfrac{1}{8} r \slashed{W} \slashed{W}^* \G^r \right) \eta_+ &=& 0, \\
\label{lastcond} \G^+ \slashed{\partial} f \psi_- + \slashed{W} \psi_-^* &=& 0. 
\eea

We can further solve (\ref{psi_plus}), by taking 
\be
\label{psi_minus}
\psi_- = \eta_- - \tfrac{1}{8} r \G^+ \slashed{W} \G^r \eta_-^*, 
\ee
where $\eta_-$ is covariantly constant, $\nabla_a^4 \eta_- = 0$.  From (\ref{proj_cond}), we then have 
\be
\G^+ \slashed{W} \eta_- = \G^+ \slashed{W}^* \eta_- = 0.  
\ee

Now, inserting (\ref{psi_minus}) into (\ref{lastcond}), we arrive at 
\be
\G^+ \left( \slashed{\partial} f - \tfrac{1}{8} r \slashed{W} \slashed{W}^* \G^r \right) \eta_- + \slashed{W} \eta^*_- = 0. 
\ee
From (\ref{EOMS}), one notes that both terms have different scalings with respect to $r$, implying that both independently vanish. Collecting the various conditions derived here, we arrive at the results quoted in the body of the paper. 

\section{Review of five-dimensional $U(1)^3$ gauged supergravity} 
In this section we present a short review of  $U(1)^3$ gauged supergravity in five dimensions. The action can be found in \cite{Cvetic:1999xp}
\bea
\label{U13theory}
\mathcal{L}_5 &=& R * \mathbf{1} - \tfrac{1}{2} \sum_{i}^2 \dd \varphi_i \wedge * \dd \varphi_i  - \tfrac{1}{2} \sum_i^3 X_i^{-2} F^i \wedge * F^i \nn 
&+& 4 g^2 \sum_i^3 X_i^{-1} \vol_5 
+ F^1 \wedge F^2 \wedge A^3,
\eea
where $g$ is a coupling constant, $F^i = \dd A^i$,  and we have defined the following scalars
\be
\label{x}
X_1 = e^{-\frac{1}{2} \left( \frac{2}{\sqrt{6}} \varphi_1 + \sqrt{2} \varphi_2 \right) }, \quad X_2 = e^{-\frac{1}{2} \left( \frac{2}{\sqrt{6}} \varphi_1 - \sqrt{2} \varphi_2 \right) }, \quad X_3 = e^{ \frac{2}{\sqrt{6}} \varphi_1 }.
\ee
Note the scalars $X_i$ are subject to the constraint $X_1 X_2 X_3 =1$. The potential possesses a single AdS$_5$ vacuum, which is supersymmetric. 

Varying the action one finds the following equations of motion:
\bea
\label{fluxeom}
\dd (X_i^{-2} * F^i) &=& \tfrac{1}{2} | \e_{ijk} | F^j \wedge F^k, 
\eea
where $i, j = 1,2, 3$ and there is no sum over $i$ on the LHS. The scalar equations of motion can be written in terms of $\varphi_i$ as 
\bea
\label{scalar_eom}
\dd * \dd \varphi_1 &=&   \tfrac{1}{\sqrt{6}} \left( X_1^{-2} F^1 \wedge * F^1 +  X_2^{-2} F^2 \wedge * F^2 - 2  X_3^{-2} F^3 \wedge * F^3   \right)  \nn 
&-& g^2 \tfrac{4}{\sqrt{6}} \left( X_1^{-1} + X_2^{-1} - 2 X_3^{-1}\right) \vol_5, \\
\dd * \dd \varphi_2 &=& \tfrac{1}{\sqrt{2}} \left( X_1^{-2} F^1 \wedge * F^1 -  X_2^{-2} F^2 \wedge * F^2  \right) - g^2 2 \sqrt{2} \left( X_1^{-1} - X_2^{-1} \right) \vol_5. \nonumber
\eea

For completeness, we also record the Einstein equation 
\bea
\label{U13Einstein} 
R_{\mu \nu} &=& \tfrac{1}{2} \sum_{i}^2 \partial_{\mu} \varphi_i \partial_{\nu} \varphi_i + \tfrac{1}{2} \sum_{i}^3 X_i^{-2} \left( F^i_{\mu \rho} F^{i~\rho}_{\nu} - \tfrac{1}{6} g_{\mu \nu} F^i_{\rho \sigma} F^{i \, \rho \sigma} \right) \nn
&-& g_{\mu \nu} \tfrac{4}{3} g^2 \sum_i^3 X_i^{-1}. 
\eea

\subsection*{Explicit five-form from KK reduction} 
Here we give an explicit form for the five-form flux including the Hodge dual. In doing so one has to consider Hodge duals for the two-dimensional space 
\be
\dd s^2 = \Delta^{-\frac{1}{2}} \frac{1}{X_i}\sum_i \dd \mu_i^2,  
\ee
for which the following expressions are useful: 
\bea
*_{2} \mathbf{1} &=&  \vol(S^2) = \tfrac{1}{2} | \e_{ijk}| \mu_i \dd \mu^j \wedge \dd \mu^k, \\
*_{2} \dd \mu_i &=& \e_{ijk} \frac{X_i}{\Delta^{\frac{1}{2}}} X_j \mu_j \dd \mu_k. 
\eea
Performing the Hodge duality, one can work out an explicit expression for the self-dual five-form 
\bea
F_5 &=& \sum_i \biggl[ 2 X_i (X_i \mu_i^2 - \Delta ) \vol_5 + \frac{1}{2} X_i^{-2} \dd (\mu^2_i) \wedge \biggl( ( \dd \varphi_i + A^i) \wedge *_5 F^i + X^i *_5 \dd X^i \biggr) \nn
&+& \left[ \sum_i 2 X_i (X_i \mu_i^2 -\Delta) \right] \frac{\mu_1 \mu_2 \mu_3}{\Delta^2} \vol(S^2) \wedge (\dd \varphi_1 + A^1)\wedge (\dd \varphi_2 + A^2)\wedge (\dd \varphi_3 + A^3) \nn
&+& F_1 \frac{\mu_2 \mu_3}{\Delta} (X_2 \mu_2 \dd \mu_3 - X_3 \mu_3 \dd \mu_2) \wedge (\dd \varphi_2 + A^2) \wedge (\dd \varphi_3 + A^3)\nn
&+& F_2 \frac{\mu_1 \mu_3}{\Delta} (X_3 \mu_3 \dd \mu_1 - X_1 \mu_1 \dd \mu_3) \wedge (\dd \varphi_3 + A^3) \wedge (\dd \varphi_1 + A^1)\nn
&+& F_3 \frac{\mu_1 \mu_2}{\Delta} (X_1 \mu_1 \dd \mu_2 - X_2 \mu_2 \dd \mu_1) \wedge (\dd \varphi_1 + A^1) \wedge (\dd \varphi_2 + A^2)\nn
&+& \frac{\mu_1 \mu_2 \mu_3}{\Delta^2} \biggl[\mu_1 \dd X_1 \wedge (X_2 \mu_2 \dd \mu_3 - X_3 \mu_3 \dd \mu_2) + \mu_2 \dd X_2 \wedge (X_3 \mu_3 \dd \mu_1 - X_1 \mu_1 \dd \mu_3) \nn &+& \mu_3 \dd X_3 \wedge (X_1 \mu_1 \dd \mu_2 - X_2 \mu_2 \dd \mu_1) \biggr] (\dd \varphi_1 + A^1) \wedge (\dd \varphi_2 + A^2) \wedge (\dd \varphi_3 + A^3).
\eea
The Bianchi identity reproduces the equations of motion (\ref{fluxeom}) and (\ref{scalar_eom}). 

\section{Some technical details} 
In this appendix we gather some technical details. 

\subsection*{Einstein equation} 
To aid the checking of the Einstein equation for the TsT of the general case (\ref{gensol1}), we here record some results. We consider  a ten-dimensional spacetime, defined by the following choice of frame 
\be
e^{+} = \tfrac{1}{r} e^{A} \dd x^+, \quad e^{-} = \tfrac{1}{r} e^{A} \left( \dd x^- - \tfrac{1}{2 f r^2}  \dd x^+ \right), \quad e^{r} =  \tfrac{1}{r} e^{A} \dd r, \quad e^{i} = \bar{e}^i, 
\ee
where $i = 1, \dots, 7$ ranges over the space transverse to the null-warped AdS$_3$ spacetime. We assume $A, f$ are independent of $r$ and that they only depend on the coordinates of the transverse space. 

The deformations we consider in section \ref{sec:gen_case} only affect the $E_{++}$ component of the Einstein equation, the corresponding Ricci tensor for which may be expressed as
\be
R_{++} = r^{-2} \left[ \tfrac{1}{2} \nabla^2 f + \tfrac{3}{2} \partial_i f \partial^i A + {4} f e^{-2A} \right].  
\ee
In calculating the Laplacian, etc, in this expression, the following orthonormal frame for the internal seven-dimensional space may be useful: 
\bea
\label{7d_ortho}
e^{3} &=& \Delta^{\frac{1}{4}} e^{g} \frac{2}{(1+ \kappa \rho^2)} \dd \rho , \nn
e^{4} &=& \Delta^{\frac{1}{4}} e^{g} \frac{2}{(1+ \kappa \rho^2)} \rho \dd \beta , \nn
e^{\mu_1} &=& \frac{ X_2^{\frac{1}{2}}  \sqrt{\Delta - X_2 \mu_2^2}}{\Delta^{\frac{1}{4}} \mu_3} \left( \dd \mu_1 + \frac{\mu_1 \mu_2 X_1}{(\Delta - X_2 \mu_2^2)} \dd \mu_2\right), \nn
e^{\mu_2} &=& \frac{\Delta^{\frac{1}{4}}}{ X_2^{\frac{1}{2}}  \sqrt{\Delta - X_2 \mu_2^2} } \dd \mu_2, \nn
e^{\varphi_i} &=& \Delta^{-\frac{1}{4}} X_i^{-\frac{1}{2}} \mu_i ( \dd \varphi_i +A^i), 
\eea
where we have opted to rewrite the metric on the internal $S^2$ in terms of $\mu_1, \mu_2$ by eliminating $\mu_3 = \sqrt{1- \mu_1^2 - \mu_2^2}$. 

\subsection*{Supersymmetry} 
Using the above orthonormal frame (\ref{7d_ortho}), we give the explicit form of the supersymmetry variation (\ref{susy_var}). The vanishing of this term leads to the condition 
\bea
\label{susy_matrix}
&& \biggl[  e^{f -2g} \frac{ \Delta^{-\frac{1}{4}} \mu_1 \mu_3 }{X_2^{\frac{1}{2}} \sqrt{\Delta - X_2 \mu_2^2}} \left( \frac{\lambda_1 a_1}{X_1^2} - \frac{\lambda_3 a_3}{X_3^2}  \right) \G^{r \mu_1} \sigma^1 \nn &&+ e^{f-2g} \frac{\m_2 X_2^{\frac{1}{2}} }{\Delta^{\frac{3}{4}} \sqrt{\Delta - X_2 \mu_2^2}} \left[ X_1 \mu_1^2 \left( \frac{\lambda_2 a_2}{X_2^2} - \frac{\lambda_1 a_1}{X_1^2} \right) + X_3 \mu_3^2 \left( \frac{\lambda_2 a_2}{X_2^2} - \frac{\lambda_3 a_3}{X_3^2} \right) \right] \G^{r \mu_2} \sigma^1 \nn 
&&+ \sum^3_{i=1} \left( \frac{2 \lambda_i \mu_i}{X_i^{\frac{1}{2}} \Delta^{\frac{1}{4}}} \G^{r \varphi_i} \sigma^3 + e^{f -2 g} \frac{\lambda_i a_i \mu_i^2}{X_i \Delta^{\frac{3}{4}}} \G^{34} \sigma^3 \right) \nn
&&-2 e^{f} \frac{\lambda_1 \mu_3 \Delta^{\frac{1}{4}} }{X_1^{\frac{1}{2}} X_{2}^{\frac{1}{2}} \sqrt{\Delta - X_2 \mu_2^2}}  \G^{\mu_1 \varphi_1} i \sigma^3 + 2 e^{f} \frac{\lambda_1 \mu_1 \mu_2 X_1^{\frac{1}{2} } X_2^{\frac{1}{2}} }{\Delta^{\frac{1}{4}} \sqrt{\Delta - X_2 \mu_2^2}} \G^{\mu_2 \varphi_1} i \sigma^3 - 2 e^{f} \frac{ \lambda_2 \sqrt{\Delta - X_2 \mu_2^2}}{\Delta^{\frac{1}{4}}} \G^{\mu_2 \varphi_2} \sigma^3 \nn 
&&+ 2 e^{f} \frac{\lambda_3 \mu_1 \Delta^{\frac{1}{4}} }{X_2^{\frac{1}{2}} X_3^{\frac{1}{2}} \sqrt{\Delta - X_2 \mu_2^2}} \G^{\mu_1 \varphi_3} \sigma^3 + 2 e^{f} \frac{\lambda_3 \mu_2 \mu_3 X_2^{\frac{1}{2}} X_3^{\frac{1}{2}} }{\Delta^{\frac{1}{4}} \sqrt{\Delta - X_2 \mu_2^2}} \G^{\mu_2 \varphi_3} \sigma^3
\biggr] \eta  = 0. 
\eea

\section{Supertube solutions} 
The purpose of this appendix is simply to remark that three-dimensional Schr\"{o}dinger solutions with dynamical exponent $z=4$ appear at the near-horizon of a class of supertube geometries. Key to this observation will be the presence of D4-brane charge and Lifshitz solutions with $z=4$ can be found in the literature \cite{Dey:2012rs}, indicating that this particular exponent, i.e. $z=4$, may be a fairly generic feature of D4-brane solutions. 

In eleven dimensions, supergravity solutions describing supertubes take the form 
\bea
\dd s^2_{11} &=& \dd s^2_5 + \left( \frac{Z_2 Z_3}{Z_1^2} \right)^{\frac{1}{3}} (\dd x_5^2 + \dd x_6^2) + \left( \frac{Z_1 Z_3}{Z_2^2} \right)^{\frac{1}{3}} (\dd x_7^2 + \dd x_8^2) +  \left( \frac{Z_1 Z_2}{Z_3^2} \right)^{\frac{1}{3}}  (\dd x_9^2 + \dd x_{10}^2), \nn
A^{(3)} &=& A^1 \wedge \dd x_5 \wedge \dd x_6 + A^2 \wedge \dd x_7 \wedge \dd x_8 + A^3 \wedge \dd x_9 \wedge \dd x_{10}, 
\eea 
where the five-dimensional spacetime can be further written as
\be
\dd s^2_5 = - (Z_1 Z_2 Z_3)^{-\frac{2}{3}} (\dd t + k)^2 +  (Z_1 Z_2 Z_3)^{\frac{1}{3}} \dd s^2_{B}. 
\ee

For the moment we will not worry about the explicit expressions,, which may be found in \cite{Bena:2007kg},  but would like to dimensionally reduce and T-dualise to recast the general solution in terms of a type IIB supergravity solution (string frame). The end result is 
\bea
\dd s^2_{IIB} &=& \left( \frac{Z_1 Z_2}{Z_3^2} \right)^{\frac{1}{6}} \dd {s}^2_5
+ \sqrt{\frac{Z_2}{Z_1}} (\dd x_5^2 + \dd x_6^2) +  \sqrt{\frac{Z_1}{Z_2}} (\dd x_7^2 + \dd x_8^2) + \frac{Z_3}{\sqrt{Z_1 Z_2}} ( \dd x_9 + A_3)^2, \nn
F^{(5)} &=& (1+*) \left[ \dd A_1 \wedge \dd x_5 \wedge \dd x_6 + \dd A_2 \wedge \dd x_7 \wedge \dd x_8 \right]  (\dd x_9 + A_3), 
\eea
We can now T-dualise along $(x_7, x_8)$ to get the solution 
\bea
\dd s^2 &=&  \left( \frac{Z_1 Z_2}{Z_3^2} \right)^{\frac{1}{6}} \dd {s}^2_5 + \sqrt{\frac{Z_2}{Z_1}} \dd s^2 (T^4) + \tfrac{1}{4} \frac{Z_3}{\sqrt{Z_1 Z_2}} ( \dd \psi + 2 A_3)^2,  \nn
F^{(3)} &=& \tfrac{1}{2} \dd A_2 \wedge (\dd \psi + 2 A_3) + \left( \frac{Z_1^2 }{Z_2 Z_3} \right)^{\frac{2}{3}} *_5 \dd A_1,   \nn
e^{2 \Phi} &=& \frac{Z_2}{Z_1}, 
\eea
where we have re-labelled $\psi = 2 x_9$ and $*_5$ refers to Hodge duality with respect to the metric $\dd s^2_5$. 

We can now compare this directly to the reduction Ansatz considered in \cite{Karndumri:2013dca} (which is based on \cite{Detournay:2012dz}) leading to the three-dimensional Einstein frame metric  
\bea
\dd s^2_3 = 4 r^4 \left[ -(\dd t + \mu \dd x)^2 + Z_1 Z_2 Z_3 (\dd x^2 + \dd r^2) \right]. 
\eea
At this stage it is a good point to introduce the explicit expressions for $\mu$, $Z_i$
\bea
Z_i = \frac{1}{2} |\e_{ijk}| K^j K^k + L_i, \quad \mu = \tfrac{1}{6} |\e_{ijk} | K^i K^j K^k + \tfrac{1}{2} K^i L_i + M, 
\eea 
where $K^i, L^i, M$ are harmonic functions, i.e. of the form $H(r) = a+ \frac{b}{r}$, where $a,b$ are constants.  

For the Ansatz considered in \cite{Colgain:2010rg} we noted that non-relativistic solutions with $z=4$ appear, but we recognise here that this is also the case in greater generality. In the limit of small $r$, we have 
\be
\mu \sim \frac{k_1 k_2 k_3}{r^3}, \quad Z^i \sim \tfrac{1}{2} |\e_{ijk} | \frac{k^j k^k}{r^2}. 
\ee

\end{document}